\documentclass[
aps,
pra,
numbers,
reprint,
nofootinbib,
longbibliography
]{revtex4-2}

\usepackage[utf8]{inputenc}

\usepackage{amsmath, amssymb, amsthm, braket, bbold}
\usepackage{graphicx, booktabs}
\usepackage{xspace}

\usepackage[caption=false]{subfig}
\usepackage{float}
\usepackage{placeins}

\newcommand{\QSL}{\textrm{QSL}\xspace}
\newcommand{\MT}{\textrm{MT}\xspace}
\newcommand{\DL}{\textrm{DL}\xspace}
\newcommand{\CZ}{\textrm{CZ}\xspace}
\newcommand{\NV}{\textrm{NV}\xspace}
\newcommand{\CPTP}{\textrm{CPTP}\xspace}
\newcommand{\POVM}{\textrm{POVM}\xspace}

\newcommand{\Lnormarrow}[2]{\ell^{#1,\max}_{#2}}
\newcommand{\Lnorm}[2]{\ell^{#1}_{#2}}
\newcommand{\FuncLnorm}[2]{L^{#1}_{#2}}
\newcommand{\Vecfun}[2][]{\VecOp_{#1}(#2)}

\newcommand{\Dnormarrowbasis}[3]{\mathcal{D}_{#1,#2,#3}^{\max}}

\newcommand{\Dnormbasis}[3]{\mathcal{D}_{#1,#2,#3}}
\newcommand{\Symnormarrow}[3]{\| #1 \|_{#2,#3}^{\max}}
\newcommand{\Symnorm}[3]{\| #1 \|_{#2,#3}}
\newcommand{\Bures}{\Theta_\text{Bures}}
\newcommand{\slbound}[2][]{\tau_{#2}^{#1}}

\DeclareMathOperator{\VecOp}{vec}
\DeclareMathOperator{\transpose}{\mathsf{T}}
\DeclareMathOperator{\Tr}{Tr}
\DeclareMathOperator{\sgn}{sgn}

\newtheorem{Thrm}{Theorem}

\newcommand{\tabledesc}[1]{\parbox[t]{0.725\columnwidth}{\raggedright #1}}

\allowdisplaybreaks

\begin{document}
	\title{Quantum Speed Limits For Open System Dynamics Based On A Representation-Basis-Dependent $\boldsymbol{\Lnorm{p}{w}}$-Seminorm}
	
	\author{H. F. Chau}
	\email{hfchau@hku.hk}
	\thanks{both authors contributed equally}
	
	\author{Jinjie Li}
	\email{jinjieli@connect.hku.hk}
	\thanks{both authors contributed equally}
	
	\affiliation{Department of Physics, University of Hong Kong, Pokfulam Road, Hong Kong}
	
	\begin{abstract}
	 We report a family of quantum speed limits (\QSL{s}) that give evolution time lower bounds between an initial and a final state whose separation is described by a certain representation basis dependent norm derived from the weighted $\Lnorm{p}{w}$-seminorm.  These \QSL{s} are applicable to open, closed, time-dependent, or time-independent systems in finite-dimensional Hilbert spaces whose density matrices are piecewise time differentiable.  They can be extended to systems over separable Hilbert spaces as well.  Crucially, these \QSL{s} are valid for arbitrary operators, not just density matrices, provided that a modest technical condition is fulfilled.  When compared to the existing \QSL{s} applied to pure state time-independent Hamiltonian evolution, qubit spontaneous emission, high-fidelity gate implementation, coherent state photon loss and operator coherence or dephasing, ours consistently show improved sharpness in most cases, along with greater universality and still retaining computational efficiency.
	\end{abstract}
	
	\keywords{$\Lnorm{p}$-norm, quantum speed limit, open system dynamics, operator coherence, dephasing dynamics}
	\maketitle

	\section{Introduction}
	\label{Sec:Intro}

	Quantum speed limit (\QSL) studies the fundamental limits on the minimum evolution time between two states. Originated from the so-called time-energy uncertainty relation, it has developed from a theoretical curiosity inquiry to a useful tool for the understanding and engineering of quantum technologies. The first \QSL was found by Mandelstam and Tamm in 1945.  This so-called \MT bound says that evolution time between two states under the action of a time-independent Hamiltonian is lower-bounded by the reciprocal of the standard deviation of the system energy~\cite{Mandelstam45}.
Interestingly, the \MT bound can be generalized to general time-dependent open system dynamics.  It reads~\cite{Braunstein96}
\begin{equation}
 \tau \ge \slbound{\MT} \equiv \frac{\hbar \Bures(\rho_\tau,\rho_0)}{\displaystyle \frac{1}{\tau} \int_0^\tau \Delta E_t \ dt} ,
 \label{E:MT}
\end{equation}
where
$\Bures(\rho_\tau,\rho_0) = \cos^{-1} \left[ \text{Tr} \left( \sqrt{\sqrt{\rho_0}\rho_\tau\sqrt{\rho_0}} \right) \right]$
is the Bures angle between the initial state $\rho_0$ and final state $\rho_\tau$, and $\Delta E_t$ is the standard deviation of the state energy at time $t$.  Here the Bures angle is the only parameter used to describe the separation between the initial and final states.  From now on, we simply refer to the time-dependent generalization bound $\slbound{\MT}$ as the \MT bound.
In 1998, Margolus and Levitin discovered a complementary \QSL for time-independent Hamiltonian evolution.  This evolution time lower bound is proportional to the reciprocal of the system's mean energy relative to the ground state~\cite{Margolus98}.
	
 \QSL research has been progressing rapidly in the last two decades.  Theoretical works have extended \QSL from its original formulation in closed, time-independent systems to include more general cases. They include open quantum systems in which interaction with the environment is non-negligible~\cite{Deffner13,Deffner17a,PDG25} as well as time-varying driven systems~\cite{Pfeifer95}. Such generalizations are important to study the ultimate evolution time limits in realistic quantum devices used in quantum information processing.
	
	An important research direction is to find more powerful and general \QSL bounds. Some of them involve geometric ideas such as the Bures angle~\cite{Anandan90, Pires16}.  Others use information theory concepts like quantum Fisher information~\cite{Taddei13, Pires16} and generalized Renyi entropy~\cite{SP25}.  While many \QSL{s} are expressed in terms of moments of the expectation values of energies of the evolving state over time, these are not the only possibilities.  For instance, \QSL bounds involving coherence~\cite{MDP22}, generalized Renyi relative entropy~\cite{SP25} and Kirkwood-Dirac quasiprobabilities~\cite{PDG25} have been reported.  Additionally, the relation between \QSL and quantum correlations has been the focus of many studies, with results showing how entanglement and other non-classical correlations can affect the speed of quantum evolution~\cite{BCPP06, Fiderer18, Maleki23}. In parallel, experimental groups have achieved significant progress in the verification of \QSL{s} in different physical platforms, such as superconducting circuits~\cite{Ness21}, NMR~\cite{Pires24}, and cavity electrodynamic systems~\cite{Cimmarusti15}. These experiments confirm the validity of the fundamental bounds and enable their application for optimizing quantum protocols.
	
	\QSL has many important implications. For quantum computing, it gives a fundamental limit for the speed of quantum gates and algorithms~\cite{Caneva09}. For quantum metrology, it limits the precision of quantum sensors~\cite{Giovannetti04, Taddei13}. Furthermore, \QSL is now considered a key concept in quantum thermodynamics, and gives insight into the efficiency and power of quantum engines~\cite{Deffner17}. As we continue to advance the frontiers of quantum science, a deep understanding of the \QSL will be necessary to utilize the full potential of the quantum realm.
	
	In this work, we introduce several \QSL{s} applicable to arbitrary quantum dynamics --- including open and time-dependent systems --- and extend their applicability beyond density matrices to any operators in separable Hilbert space as long as certain technical conditions are fulfilled. Our main contributions are:
	
	\begin{itemize}
	 \item \textbf{Unified \QSL framework}: We introduce a representation basis dependent weighted $\Lnormarrow{p}{w}$-norm derived from the weighted $\Lnorm{p}{w}$-seminorm, and use it to obtain three closely related \QSL{s} for finite dimensional systems that are applicable for both open and closed dynamics as long as the density matrices are piecewise time differentiable.  We further extend this formalism to countably infinite dimensional systems.  More importantly, our framework generalizes traditional state-based \QSL{s} to linear operators.  Unlike most \QSL{s} on market, ours are representation basis dependent in general.  In fact, our \QSL{s} are neither based on distinguishability measure nor monotone under \CPTP maps.
	 \item  \textbf{Efficient and numerically stable computation}:
	  For finite dimensional systems, two of our \QSL{s} can be computed efficiently in a numerically stable manner.  Besides, our remaining \QSL, which requires optimization over the representation basis, can be efficiently computed in many cases of practical interest.  Nevertheless, just like a generic optimization problem, we do not see a general pattern of the optimal representation basis that gives us the best \QSL.  All these comments apply also to computing our \QSL{s} for countably infinite dimensional systems that can be uniformly approximated by discretization or truncation.
	 \item \textbf{Demonstration of wide applicability}: We illustrate the wide applicability of our \QSL{s} by showcasing their use in the following five systems.
		 \begin{itemize}
		  \item Closed time-independent Hamiltonian evolution of certain two-dimensional qubit and four-dimensional qudit system that saturate one of our \QSL bounds;
			\item Spontaneous emission in a qubit, benchmarking against the Deffner-Lutz (\DL) bound~\cite{Deffner13};
			\item High-fidelity gate operations in nitrogen-vacancy (\NV) centers in diamond, comparing robustness against the \MT-bound \cite{Herb24};
			\item Photon loss of a coherent state, demonstrating the applicability and superiority of our bound over the \DL bound~\cite{Deffner13} for countably infinite dimensional quantum systems; and
			\item Dephasing and coherence generation for general Hilbert space operators, illustrating the formalism's extensibility beyond density matrices \cite{Shrimali25}.
		\end{itemize}
	\end{itemize}

	Just like most \QSL{s} on market (Ref.~\cite{Chau13} is a notable exception), ours are based on a single parameter describing the separation between the initial and final states.  We defer the study of \QSL in the multiple parameters situation to our subsequent work.

	We begin by stating our motivation and intuition behind before introducing our \QSL{s} and their proofs in Sec.~\ref{Sec:QSL}.  Then we illustrate the use of these \QSL{s} using five examples covering time-independent, time-dependent, finite dimensional and countably infinite dimensional system evolution as well as general Hilbert space operator \QSL{s} in Sec.~\ref{Sec:Appl}.  Finally, we summarize our findings and point out a few meaningful future research directions in Sec.~\ref{Sec:Discussion}.

	\section{Quantum Speed Limits Based On A Representation-Dependent Weighted $\boldsymbol{\Lnorm{p}{w}}$-Seminorm}
	\label{Sec:QSL}
	\subsection{Main Results And Proofs}
	\label{Subsec:Proof}
We wanted to find efficiently and accurately computable \QSL{s} that are more powerful than existing ones for a wide range of finite dimensional systems including time-dependent and open ones.  Besides, it would be nice if these \QSL{s} can be generalized to infinite dimensional systems.  To achieve these goals, we need a density matrix distance measure or equivalently a norm that make use of as much information on the density matrices of the system as possible.  Based on this demand, we introduce the following norm for finite dimensional systems.  And we are going to discuss its generalization to countably infinite dimensional systems later in this Section.  Since a number of symbols and notations are needed to better articulate our formulation and analysis, we summarize them in Table~\ref{T:notations} for easy reference.

	Let $\rho_\tau \equiv \rho(\tau)$ be the density matrix of a $n$-dimensional quantum state as a function of time variable $\tau$.  In a fixed orthonormal basis $\mathcal{B}$ representation, its matrix elements can be written as a $n^2$-dimensional vector through the following column-stacking vectorization~\cite{Watrous}
	\begin{equation}
	 \Vecfun[\mathcal{B}]{\rho(\tau)} = (\rho(\tau)_{11},\rho(\tau)_{12},\cdots,\rho(\tau)_{nn}) \in \mathbb{C}^{n^2} .
	 \label{E:Vecfun_def}
	\end{equation}
	We further denote time-dependent difference vector by
	\begin{equation}
	 \Delta_\mathcal{B}(\tau) = \Vecfun[\mathcal{B}]{\rho_\tau} - \Vecfun[\mathcal{B}]{\rho_0} .
	 \label{E:time_diff_def}
	\end{equation}
	(We shall omit the basis subscript $\mathcal{B}$ in $\VecOp$ and $\Delta$ if it is either understood from the context or unimportant.)
	Lastly, for any fixed $p \ge 1$ and non-zero real vector $w \in \mathbb{R}^{n^2}$ whose elements are non-negative, we equip the set of all $n^2$-dimensional vectors (and hence also the set of all density matrices) with the $\Lnormarrow{p}{w}$-norm, or more precisely the $\Lnormarrow{p}{w}([n^2])$-norm with $[n^2]$ denotes the set $\{ 0,1,2,\cdots,n^2-1 \}$, where
	\begin{subequations}
	 \label{E:norm_def}
	\begin{align}
	 & \Dnormarrowbasis{p}{w}{\mathcal{B}}(\rho_\tau, \rho_0) \notag \\
	 ={}& \Symnormarrow{\Delta_\mathcal{B}(\tau)}{p}{w} \equiv \max_{P\in S_{n^2}} \left[ \sum_{j=1}^{n^2} w_{P(j)} |\Delta_{\mathcal{B},j}(\tau)|^p \right]^{1/p} \label{E:norm_def1} \\
	 ={}& \left\{ \sum_{j=1}^{n^2} w^\downarrow_j \left[ |\Delta_\mathcal{B}(\tau)|_j^\downarrow \right]^p  \right\}^{1/p} \label{E:norm_def2} .
	\end{align}
	\end{subequations}
	Here $S_{n^2}$ is the symmetric group on $n^2$ elements and for any real vector $v$, $v^\downarrow_j$ refers to the $j$th resultant element by arranging elements of $v$ in descending order.  Besides, Corollary~II.4.4 in Ref.~\cite{BhatiaMatrixAnalysis} is used to obtain Eq.~\eqref{E:norm_def2} from Eq.~\eqref{E:norm_def1}.  Since $\Lnorm{p}{w}{([n^2])}$ is a seminorm for all $w$, our construction in Eq.~\eqref{E:norm_def1} automatically makes $\Symnormarrow{\cdot}{p}{w}$ a seminorm.  But actually, $\Symnormarrow{\cdot}{p}{w}$ is a norm in case some of the elements in $w$ are $0$.  The reason is that if $\Symnormarrow{\Delta_\mathcal{B}(\tau)}{p}{w} = 0$, Eq.~\eqref{E:norm_def2} and the fact that $w$ is a non-zero vector whose elements are non-negative imply that $w^\downarrow_1 > 0$ and $|\Delta_\mathcal{B}(\tau)|_1^\downarrow = 0$.  Thus, $|\Delta_\mathcal{B}(\tau)|_j^\downarrow = 0$ for all $j$.  Consequently, $\Delta_\mathcal{B}(\tau) = 0$.  Last but not least, by continuity, we define $\Dnormarrowbasis{\infty}{w}{\mathcal{B}}(\rho_t,\rho_0)$ as $\Symnormarrow{\Delta_\mathcal{B}(\tau)}{\infty}{w} = |\Delta_\mathcal{B}(\tau)|_1^\downarrow$.  Readers may refer to the two examples in Secs.~\ref{Subsec:Appl_Pure_State} and~\ref{Subsec:Appl_SpontEmission} on explicit computation of $\Dnormarrowbasis{p}{w}{\mathcal{B}}(\cdot,\cdot)$ in two simple cases.

\begin{table}
 \begin{tabular}{ll}
   \toprule
   Notation & Meaning
  \\
   \midrule
   $\rho_t, \rho(t)$ & \tabledesc{Density matrix at time $t$}
  \\
   $\mathbb{L}$ & \tabledesc{Lindblad superoperator}
  \\
   $p$, $q$ & \tabledesc{Hölder conjugate pair}
  \\
   $w$ & \tabledesc{Non-zero weight vector}
  \\
   $\bar{w}$ & \tabledesc{reordered version of $w$ defined in Eq.~\eqref{E:bar_w_def}}
  \\
   $\mathcal{B}$ & \tabledesc{Orthonormal basis}
  \\
   $\mathcal{B}_c, \mathcal{B}_d, \mathcal{B}_F, \mathcal{B}_E$ & \tabledesc{Computational, diagonal, Fock bases and energy eigenbasis}
  \\
   $v^\downarrow_j$ & \tabledesc{The $j$th element of a vector $v$ arranged in descending order}
  \\
   $s^\downarrow_j(A)$ & \tabledesc{The $j$th singular value of $A$ arranged in descending order}
  \\
   $\mathbb{1}$ & \tabledesc{The vector whose elements are all $1$}
  \\
   $\mathbb{1}_j$ & \tabledesc{The vector those first $j$ elements are $1$ and the rest are $0$}
  \\
   $S_{n^2}, S_{\mathbb{N}}$ & \tabledesc{Symmetric groups of $[n^2]$ and $\mathbb{N}$}
  \\
   $\Vecfun[]{\cdot}, \Vecfun[\mathcal{B}]{\cdot}$ & \tabledesc{Column-stacking vectorization defined in Eq.~\eqref{E:Vecfun_def}}
  \\
   $\Delta(\tau), \Delta_\mathcal{B}(\tau)$ & \tabledesc{Time-dependent difference vector defined in Eq.~\eqref{E:time_diff_def}}
  \\
   $\Dnormarrowbasis{p}{w}{\mathcal{B}}(\cdot,\cdot)$ & \tabledesc{Metric defined in Eqs.~\eqref{E:norm_def} and~\eqref{E:norm_extended_def}}
  \\
   $\Symnormarrow{\cdot}{p}{w}$ & \tabledesc{$\Lnormarrow{p}{w}([n^2])$- or $\Lnormarrow{p}{w}(\mathbb{N})$-norm defined in Eqs.~\eqref{E:norm_def} and~\eqref{E:norm_extended_def}}
  \\
   $\Dnormbasis{p}{\bar{w}}{\mathcal{B}}(\cdot,\cdot)$ & \tabledesc{Pseudometric defined in Eq.~\eqref{E:bar_w_def}}
  \\
   $\Symnorm{\cdot}{p}{\bar{w}}$ & \tabledesc{Seminorm defined in Eq.~\eqref{E:bar_w_def}}
  \\
   $\slbound{\CZ}, \slbound{\DL}, \slbound{\MT}$ & \tabledesc{\CZ, \DL, and \MT \QSL bounds}
  \\
   $\slbound[\text{int}]{}, \slbound[\text{int}]{p,w^\downarrow,\mathcal{B}}$ & \tabledesc{Basic integral form \QSL in Eq.~\eqref{E:new_QSL_integral}}
  \\
   $\slbound[\text{sup}]{}, \slbound[\text{sup}]{p,w^\downarrow,\mathcal{B}}$ & \tabledesc{Basic supremum form \QSL in Eq.~\eqref{E:new_QSL}}
  \\
   $\slbound[\text{int}]{\text{opt}}$ & \tabledesc{Optimized form \QSL}
  \\
   \bottomrule
 \end{tabular}
 \caption{List of frequently used notations.
  \label{T:notations}}
\end{table}

	Our intuition is that the norm property of $\Dnormarrowbasis{p}{w}{\mathcal{B}}$ demands $\displaystyle \Dnormarrowbasis{p}{w}{\mathcal{B}}(\rho_\tau,\rho_0) \le \int_0^\tau \left| \frac{d \Dnormarrowbasis{p}{w}{\mathcal{B}}}{dt} \right| dt$.  So, a upper bound for the integrand gives a \QSL for the system.  The smaller the bound, the more powerful the \QSL.  Nevertheless, this method is tricky in actual numerical calculation because the optimal $P \in S_{n^2}$ that maximizes the above integrand is time dependent.  Accurate computation requires precise determination of the time when the optimal $P$ changes.  Besides, theoretical analysis of the convergence of this integral is non-trivial.  Hence, we consider a variation of the theme.  But before further discussion, we need to introduce the following notation.

	Let $P$ be the permutation in $S_{n^2}$ that maximizes Eq.~\eqref{E:norm_def1}.  We denote $\bar{w}$ the reordered version of $w$ given by $\bar{w}_j = w_{P(j)}$ for all $j$.  Using this notation,
\begin{align}
 \Dnormarrowbasis{p}{w}{\mathcal{B}}(\rho_\tau, \rho_0) &= \left[ \sum_{j=1}^{n^2} \bar{w}_j |\Delta_{\mathcal{B},j}(\tau)|^p \right]^{1/p} \notag \\
 &\equiv \Dnormbasis{p}{\bar{w}}{\mathcal{B}}(\rho_t,\rho_0) \equiv \Symnorm{\Delta_{\mathcal{B}}(\tau)}{p}{\bar{w}} .
 \label{E:bar_w_def}
\end{align}
Surely, this permuted weight vector $\bar{w}$ depends on $p$, $w$, $\mathcal{B}$ and may vary with $\Delta_\mathcal{B}(\tau)$.  Besides, it is not unique in general.  Nonetheless, Eq.~\eqref{E:bar_w_def} holds irrespective of the choice of $P$ (and hence $\bar{w}$).

Using this notation, the seminorm property of $\Lnorm{p}{w}([n^2])$ plus Eq.~\eqref{E:norm_def2}, we have $\displaystyle \Dnormarrowbasis{p}{w}{\mathcal{B}}(\rho_\tau,\rho_0) \le \int_0^\tau \left| \frac{d \Dnormbasis{p}{\bar{w}}{\mathcal{B}}}{dt} \right| dt$.  Now the weight vector $\bar{w}$ used throughout the integral over time is fixed.  Surely, the smaller the upper bound for this integrand, the better the resultant \QSL.  This is the core concept used in our \QSL{s} stated in the Theorem below.

\begin{Thrm}
 Let $\rho_t$ be the density matrix of a finite-dimensional quantum system at time $t$.  Suppose that the evolution of $\rho_t$ is governed by the Lindblad superoperator $\mathbb{L}$.  Suppose further that $\rho_t$ is differentiable for all $t\in [0,\tau]$ except possibly at finitely many times $t_1 <  t_2 < \cdots < t_{k-1}$.  Then for any $p \in [1,\infty]$, the evolution time $\tau \ge 0$ obeys
	\begin{subequations}
	 \label{E:new_QSL_group}
	\begin{align}
	 \tau &\geq \slbound[\text{\rm int}]{p,w^\downarrow,\mathcal{B}} \equiv \frac{\Dnormarrowbasis{p}{w}{\mathcal{B}}(\rho_\tau, \rho_0)}{\displaystyle \frac{1}{\tau} \int_0^\tau \Symnorm{\!\Vecfun[\mathcal{B}]{\mathbb{L} \rho_t}}{p}{\bar{w}} \ dt} \label{E:new_QSL_integral} \\
	 &\geq \slbound[\sup]{p,w^\downarrow,\mathcal{B}} \equiv \frac{\Dnormarrowbasis{p}{w}{\mathcal{B}}(\rho_\tau, \rho_0)}{\displaystyle \max_{j = 0}^{k-1} \sup_{t\in (t_j,t_{j+1})} \Symnorm{\!\Vecfun[\mathcal{B}]{\mathbb{L} \rho_t}}{p}{\bar{w}}}
	 \label{E:new_QSL}
	\end{align}
	\end{subequations}
 where $t_0 = 0$ and $t_k = \tau$ if $\Dnormarrowbasis{p}{w}{\mathcal{B}}(\rho_\tau,\rho_0) > 0$.
 \label{Thrm:main}
\end{Thrm}
	
	\begin{proof}
	 Note that $\tau > 0$ for $\Dnormarrowbasis{p}{w}{\mathcal{B}}(\rho_\tau,\rho_0) > 0$.  Thus, $\Dnormarrowbasis{p}{w}{\mathcal{B}}(\rho_t,\rho_0)$ and $\Dnormbasis{p}{\bar{w}}{\mathcal{B}}(\rho_t,\rho_0)$ can be regarded as functions of $t\in [0,\tau]$.  As $\rho_t$ satisfies the Lindblad equation $\dot{\rho_t} = \mathbb{L}\rho_t$ and $\rho_t$ can be regarded as a finite-dimensional linear operator, $\Dnormarrowbasis{p}{w}{\mathcal{B}}(\rho_t,\rho_0)$ is finite for all $t\in [0,\tau]$.  Moreover, it is differentiable at $t \in J \equiv \bigcup_{j=0}^{k-1} (t_j,t_{j+1})$ by the differentiability assumption of $\rho_t$.  Clearly,
	 \begin{align}
	  & \frac{d \Dnormarrowbasis{p}{w}{\mathcal{B}}}{dt} = \frac{d \Dnormbasis{p}{\bar{w}}{\mathcal{B}}}{dt} \notag \\
	  ={}& \Dnormbasis{p}{\bar{w}}{\mathcal{B}}^{1-p} \sum_{j=1}^{n^2} \bar{w}_j |\Delta_{\mathcal{B},j}|^{p-1} \sgn(\Delta_{\mathcal{B},j}) \ \frac{d\!\Vecfun[\mathcal{B}]{\rho_t}_j}{dt} .
	  \label{E:dDdt}
	\end{align}

	 Let $p > 1$ and $1/p + 1/q = 1$.  By applying Hölder inequality to $\bar{w}_j^{1/p} \frac{d}{dt}\!\Vecfun[\mathcal{B}]{\rho_t}_j$ and $\bar{w}_j^{1/q} |\Delta_{\mathcal{B},j}|^{p-1} \sgn(\Delta_{\mathcal{B},j})$ and followed by multiplying both sides of the resultant inequality by $\Dnormbasis{p}{\bar{w}}{\mathcal{B}}^{1-p}$, we obtain
	 \begin{equation}
	\left|\frac{d\Dnormarrowbasis{p}{w}{\mathcal{B}}}{dt}\right| \leq \left\|\frac{d\!\Vecfun[\mathcal{B}]{\rho_t}}{dt}\right\|_{p,\bar{w}}
	 = \Symnorm{\!\Vecfun[\mathcal{B}]{\mathbb{L} \rho_t}}{p}{\bar{w}}
	  \label{E:intermediate}
	 \end{equation}
	 for all $t\in J$.  Here, $\Symnorm{\cdot}{p}{\bar{w}}$ is the $\Lnorm{p}{\bar{w}}([n^2])$-seminorm.  By taking limits on $p$, it is clear that this inequality holds also for the case of $p = 1$ or $\infty$.

	 Observe that $\mathbb{L}$ is a bounded superoperator acting on finite-dimensional density matrices.  Therefore, $\displaystyle \sup_{t\in J} \Symnorm{\!\Vecfun[\mathcal{B}]{\mathbb{L} \rho_t}}{p}{\bar{w}}$ is well-defined and finite.  In addition, Lebesgue dominated convergence theorem guarantees that $\displaystyle \int_0^\tau \Symnorm{\!\Vecfun[\mathcal{B}]{\mathbb{L} \rho_t}}{p}{\bar{w}} \ dt$ exists.  Furthermore, these integral and supremum must be positive for otherwise $\Dnormarrowbasis{p}{w}{\mathcal{B}}(\rho_\rho,\rho_0) = 0$.  Consequently, Inequalities~\eqref{E:new_QSL_integral} and~\eqref{E:new_QSL} are obtained by integrating Eq.~\eqref{E:intermediate} over time.  This completes our proof.
	\end{proof}

	Inequalities~\eqref{E:new_QSL_integral} and~\eqref{E:new_QSL} are our basic \QSL{s}.  We call $\slbound[\text{int}]{p,w^\downarrow,\mathcal{B}}$ or simply $\slbound[\text{int}]{}$ the basic integral form \QSL.  And we call $\slbound[\text{int}]{p,w^\downarrow,\mathcal{B}}$ or simply $\slbound[\text{sup}]{}$ the basic supremum form \QSL.
	Several remarks are in place.

	\begin{enumerate}
	 \item Theorem~\ref{Thrm:main} is still valid for the trivial case of $\Dnormarrowbasis{p}{w}{\mathcal{B}}(\rho_\tau,\rho_0) = 0$ if we interpret $\slbound[\text{int}]{p,w^\downarrow,\mathcal{B}} = \slbound[\text{sup}]{p,w^\downarrow,\mathcal{B}} = 0$ with no regard to the values of the denominators in the R.H.S. of Inequalities~\eqref{E:new_QSL_integral} and~\eqref{E:new_QSL}.
	 \item Our \QSL{s} are derived by bounding the norm $\Dnormarrowbasis{p}{w}{\mathcal{B}}(\rho_\tau,\rho_0)$ through integrating over the rate of change of the auxiliary semi-norm $\Symnorm{\!\Vecfun[\mathcal{B}]{\mathbb{L}\rho_t}}{p}{\bar{w}}$.  That is to say, our derivation make no use of the norm property of $\Dnormarrowbasis{p}{w}{\mathcal{B}}$.  So, our \QSL{s} are valid for any $\Dnormbasis{p}{w}{\mathcal{B}}$, too.  However, these extended \QSL{s} are generally less sharp.  That is why they are not studied here.
	 \item \label{Rem:basis} Changing basis in $\rho_t$ corresponds to a change of $\mathcal{B}$ in $\Dnormarrowbasis{p}{w}{\mathcal{B}}(\rho_\tau,\rho_0)$.  Since both $\Dnormarrowbasis{p}{w}{\mathcal{B}}$ and $\Symnorm{\!\Vecfun[\mathcal{B}]{\mathbb{L}\rho_t}}{p}{\bar{w}}$ depend on $\mathcal{B}$, our basic \QSL{s} in Inequality~\eqref{E:new_QSL_group} are not unitarily invariant in general.  In fact, this is what we observe in all the applications to be reported in Sec.~\ref{Sec:Appl}.  (But see Sec.~\ref{Subsubsec:Properties_p=2} for an important exception.)  Consequently, though having similar forms, our basic \QSL{s} are different from those of the \MT bound~\cite{Braunstein96} as well as the $\Lnorm{p}{}$- or $\FuncLnorm{p}{}$-norms-based ones reported in Refs.~\cite{Deffner17,Chau10,LeeChau13} for theirs are unitarily invariant.  More comparison between these \QSL{s} will be given in Sec.~\ref{Subsubsec:Properties_p=2} below.
	 \item Since unitary transformation is a special case of \CPTP map, a direct and interesting consequence of the basis-dependent property of the metric $\Dnormarrowbasis{p}{w}{\mathcal{B}}$ is that this metric is not a monotone under \CPTP map.  (But again, please refer to Sec.~\ref{Subsubsec:Properties_p=2} for an important exception.)  As a consequence, unlike the \MT bound, our metric $\Dnormarrowbasis{p}{w}{\mathcal{B}}$ together with our \QSL{s} are in general not based on distinguishability measure of quantum states.  Here we remark that non-distinguishability-based metric is not uncommon in quantum information science.  Frobenius norm is a famous example.
	 \item $\Dnormarrowbasis{p}{w}{\mathcal{B}}(\rho_t,\rho_0)$ and $\Symnorm{\!\Vecfun[\mathcal{B}]{\mathbb{L} \rho_t}}{p}{\bar{w}}$ are observables for any time $t$.  Determining them may require state tomography or \POVM measurement (on multiple and identical copies of the state of the system) at all times $t \in [0,\tau]$.  As $\rho_t$ traces out a path in the column-stacking vectorization $\Vecfun[\mathcal{B}]{\rho(t)}$, $\slbound[\text{int}]{p,w^\downarrow,\mathcal{B}}$ and $\slbound[\text{sup}]{p,w^\downarrow,\mathcal{B}}$ originate from the average and supremum rate of change of this path, respectively.  Combined with Remark~\ref{Rem:basis} above, our two non-unitary-invariant \QSL{s} offer rooms for optimization that are not present in all unitary invariant \QSL{s}.
	 \item Our \QSL{s} are unchanged when $w$ is replaced by $\lambda w$ for any $\lambda > 0$.  Thus, without loss of generality, we may fix $w_1^\downarrow = 1$.  In addition, if all but one of the components of $w$ are $0$ (that is to say, $w_j^\downarrow = 0$ for all $j > 1$), then Eq.~\eqref{E:norm_def2} implies that our \QSL{s} do not depend on $p$.
	 \item \label{Rem:tie} For given $p$, $w$, $\mathcal{B}$, and in the event that the corresponding $\bar{w}$ is not unique, then surely we pick the one that maximizes either Inequality~\eqref{E:new_QSL_integral} or~\eqref{E:new_QSL}.  Interestingly, for all the cases with non-unique $\bar{w}$ we are going to study in Sec.~\ref{Sec:Appl} below, the values of $\slbound[\text{int}]{}$ and $\slbound[\max]{}$ are does not depend on the choice of $\bar{w}$.  It is instructive to find out why in future.
	 \item \label{Rem:sup} Although $\slbound[\text{int}]{}$ generally gives a better lower bound than $\slbound[\text{sup}]{}$, finding it requires precisely known $\rho_t$ that may not be available (due to uncertainty in the initial state or the evolution model, for instance) as well as integration which can be computationally taxing sometimes. In contrast, calculating $\slbound[\text{sup}]{}$ or its lower bound is often simpler, more robust and computationally less demanding.  As we shall illustrate in Secs.~\ref{Subsec:Appl_SpontEmission} and~\ref{Subsec:Appl_NV} below, $\slbound[\text{sup}]{}$ occasionally gives good \QSL bounds.
	\end{enumerate}

	Surely, we may further optimize these basic \QSL{s} over $p$, $w$ as well as the basis $\mathcal{B}$ used to construct $\rho_t$ and $\rho_0$.  It turns out that optimizing $\slbound[\text{int}]{}$ is more useful; and we denote this optimized form \QSL by $\slbound[\text{int}]{\text{opt}}$ and refer to it as the (fully) optimized form.  If $\Symnorm{\!\VecOp(\mathbb{L} \rho_t)}{p}{\bar{w}}$ is sufficiently piecewise smooth, then optimization over $p$ can be done efficiently at least for finding local maxima.  Likewise, optimization over the basis in constructing $\Delta_\mathcal{B}(\tau)$ and $\Vecfun[\mathcal{B}]{\mathbb{L} \rho_t}$ is a variational optimization problem for smooth function on compact a Lie group that can be done efficiently at least for local solutions~\cite{Smith93,TK94,TO20}.  Finally, efficient optimization over $w$ can be found in Sec.~\ref{Subsubsec:Properties_w} below.  To summarize, local maxima for optimizing $\slbound[\text{int}]{p,w,\mathcal{B}}$ can be found efficiently.  In practice, we find that the optimized form \QSL $\slbound[\text{int}]{\text{opt}}$ can be found efficiently almost all the time.  In contrast, we do not know of any analytical way to find $\slbound[\text{int}]{}$.  In fact, readers are going to see in Sec.~\ref{Sec:Appl} below that there is no obvious pattern for the basis $\mathcal{B}$ needed in the simpler problem of maximizing $\slbound[\text{int}]{p,w^\downarrow,\mathcal{B}}(\rho_\tau,\rho_0)$ or $\slbound[\text{sup}]{p,w^\downarrow,\mathcal{B}}(\rho_\tau,\rho_0)$ given $p$ and $w^\downarrow$.

	Along a different line, we point out that our derivation does not make use of density matrix property.  In fact, our \QSL{s} are still valid if $\rho_t$ is substituted by a linear operator $A$, even a non-Hermitian one as long as the matrix elements of the operator are continuous for all $t\in [0,\tau]$, differentiable for all $t \in J$, and $\Symnorm{\!\Vecfun[\mathcal{B}]{\mathbb{L} A}}{p}{\bar{w}}$ is bounded or more generally dominated by a Lebesgue integrable function over time interval $[0,\tau]$.

Last but not least, our \QSL bounds can be extended to states evolving in countably infinite dimensional Hilbert spaces.  In this case, for a fixed basis $\mathcal{B}$, we may vectorize any linear operator in basically the same way as in Eq.~\eqref{E:Vecfun_def} by writing, for instance, the column-stacking vector for the density matrix $\sigma_\tau$ of a countably infinite dimensional Hilbert space as $\Vecfun[\mathcal{B}]{\sigma_\tau} = (\Vecfun[\mathcal{B}]{\sigma_\tau}_{j+1})_{j\in \mathbb{N}}$.  (It may be more apparent to use $\mathbb{N}^2$ as the index set for $j$ so as to emphasize the action of column-stacking matrix elements of $\sigma_\tau$.  However, we choose to use $\mathbb{N}$ which has the same cardinality of $\mathbb{N}^2$ instead because there is an natural ordering of $\mathbb{N}$ so that we could talk about $w^\downarrow_{j+1}$ for any $j\in \mathbb{N}$ without ambiguity in the same way as in Eq.~\eqref{E:norm_def2}.  Here we use $j+1$ instead of $j$ as the index so that the notation $w^\downarrow_{j+1}$ coincides with the standard usage in matrix analysis.)  Then for any non-zero vector $w = (w_{j+1})_{j\in \mathbb{N}}$ with $w_{j+1} \ge 0$, we may extend the definition in Eq.~\eqref{E:norm_def} to
\begin{subequations}
 \label{E:norm_extended_def}
 \begin{align}
  & \Dnormarrowbasis{p}{w}{\mathcal{B}}(\rho_\tau,\rho_0) \notag \\
  ={}& \Symnormarrow{\Delta_\mathcal{B}(\tau)}{p}{w} \equiv \max_{P\in S_{\mathbb{N}}} \left[ \sum_{j\in \mathbb{N}} w_{P(j)+1} |\Delta_{\mathcal{B},j+1}(\tau)|^p \right]^{1/p} \label{E:norm_extended_def1} \\
  ={}& \left\{ \sum_{j\in \mathbb{N}} w^\downarrow_{j+1} \left[ |\Delta_\mathcal{B}(\tau)|_{j+1}^\downarrow \right]^p  \right\}^{1/p} \label{E:norm_extended_def2}
 \end{align}
\end{subequations}
 whose validity in notation and as a norm will be justified in
 Appendix~\ref{Sec:norm_extension}.  Using this extended definition and denoting the element in $S_{\mathbb{N}}$ that maximizes Eq.~\eqref{E:norm_extended_def1} by $\bar{w}$, we state our \QSL{s} for countably infinite dimensional system below whose proof can be found in Appendix~\ref{Sec:proof_extension}.

\begin{Thrm}
 Theorem~\ref{Thrm:main} holds for countably infinite dimensional quantum systems provided that the following conditions are satisfied:
 \begin{enumerate}
  \item $\Dnormarrowbasis{p}{w}{\mathcal{B}}(\rho_t,\rho_0)$ is finite for $t
   \in [0,\tau]$;
   \label{Cond:finiteness}
  \item $\displaystyle \sum_{j=0}^\ell \frac{df_j(t)}{dt}$ converges uniformly
   in each of the time interval $(t_m,t_{m+1})$ as $\ell\to\infty$ for $m =
   0,1,\cdots ,k$, where $f_j(t) = w^\downarrow_{j+1} \left[
    |\Delta_{\mathcal{B}}(t)|^\downarrow_{j+1} \right]^p$;
   \label{Cond:uniform_convergence}
  \item For all $t\in J$, $\Symnorm{(\bar{w}_{j+1}^{1/p} \frac{d}{dt} \!\Vecfun[\mathcal{B}]{\rho_t}_{j+1})_{j\in \mathbb{N}}}{p}{\mathbb{1}}$ and $\Symnorm{(\bar{w}^{1/q}_{j+1} |\Delta_{\mathcal{B},j+1}(t)|^{p-1} \sgn(\Delta_{\mathcal{B},j+1}(t)))_{j\in \mathbb{N}}}{q}{\mathbb{1}}$ are finite where $\mathbb{1}$ denotes the vector whose elements are all $1$; and
   \label{Cond:Holder}
  \item $\Symnorm{\!\Vecfun[\mathcal{B}]{\mathbb{L} \rho_t}}{p}{\bar{w}} \le g(t)$ for some $g(t) \in \FuncLnorm{1}{}[0,\tau]$.
   \label{Cond:LDCT}
 \end{enumerate}
 Here, $\slbound[\text{sup}]{p,w^\downarrow,\mathcal{B}}$ is regarded as $0$ if the denominator in the R.H.S. of Eq.~\eqref{E:new_QSL} is infinite.
 \label{Thrm:extension}
\end{Thrm}

\subsection{Properties Of The Basic Bounds}
\label{Subsec:Properties}

\subsubsection{Sufficient Condition for The Bounds to be Representation Basis Independent}
\label{Subsubsec:Properties_p=2}
As pointed out in Remark~\ref{Rem:basis}, $\slbound[\text{int}]{p,w^\downarrow.\mathcal{B}}$ and $\slbound[\text{sup}]{p,w^\downarrow,\mathcal{B}}$ are representation basis dependent in general. Nevertheless, there is a notable exception.  Observe that any linear operator $A$ acting on $\mathcal{H}^n$ must satisfy $\sum_{j,k} |A_{jk}|^2 = \sum_{j,k} \bra{j} A \ket{k} \bra{k} A^\dag \ket{j} = \Tr A A^\dag = \sum_j s_j^\downarrow(A)^2$, where $s_j^\downarrow(A)$ denotes that $j$th singular value of $A$ arranged in descending order.  As a result,
\begin{equation}
 \Symnorm{\!\Vecfun[\mathcal{B}]{\mathbb{L} \rho_t}}{2}{\mathbb{1}_{n^2}} = \left[ \sum_j s_j^\downarrow(\mathbb{L} \rho_t)^2 \right]^{1/2}
 \label{E:symnorm_2_1}
\end{equation}
where $\mathbb{1}_j$ is the vector of length $n^2$ whose first $j$ components are $1$ and the remaining $(n^2 - j)$ components are $0$.  Consequently, our basic \QSL{s} in Inequality~\eqref{E:new_QSL_group} for $p = 2$ and $w = \mathbb{1}_{n^2}$ are independent of the basis used to construct $\rho_t$ and $\rho_0$.  In fact, $\Symnorm{\cdot}{2}{\mathbb{1}_{n^2}}$ equals the Frobenius norm $\| \cdot \|_2$. In addition, for any sufficiently small time interval $[t,t+\Delta t]$, $\mathbb{L} \rho_t = -i [\sum_j E_j(t) \ket{E_j(t)}\bra{E_j(t)}, \rho_t]/\hbar + \text{O}(\Delta t^2)$ where $E_j(t)$ and $\ket{E_j(t)}$ are the eigenenergy and the corresponding eigenstate of the system at time $t$.  Hence,
\begin{subequations}
\begin{equation}
 \slbound[\text{int}]{2,\mathbb{1}_{n^2}^\downarrow} = \frac{\left\| \rho_t - \rho_0 \right\|_2}{\displaystyle \frac{1}{\tau} \int_0^\tau \left\| \mathbb{L} \rho_t \right\|_2 \ dt} = \frac{\hbar \left\| \rho_t - \rho_0 \right\|_2}{\displaystyle \frac{\sqrt{2}}{\tau} \int_0^\tau \Delta E_t \ dt}
\end{equation}
and
\begin{equation}
 \slbound[\text{sup}]{2,\mathbb{1}_{n^2}^\downarrow} = \frac{\left\| \rho_t - \rho_0 \right\|_2}{\displaystyle \sup_{t\in [0,\tau]} \left\| \mathbb{L} \rho_t \right\|_2} = \frac{\hbar \left\| \rho_t - \rho_0 \right\|_2}{\displaystyle \sqrt{2} \sup_{t\in [0,\tau]} \Delta E_t} .
 \label{E:alt_QSL}
\end{equation}
\end{subequations}
In this regard, our basic integral form \QSL and the \MT bound can be expressed in almost the same form.  The difference is that our numerator depends on the Frobenius distance between the initial and final states while theirs depends on the Bures angle.  This conclusion is consistent with our observation in Remark~\ref{Rem:basis}.  Likewise, we conclude by extension that although our \QSL{s} appear to have the same form as in those in Refs.~\cite{Chau10,LeeChau13}, they are of very different nature.  This is because in the time-independent case, $\Symnorm{\!\Vecfun[\mathcal{B}]{\mathbb{L}\rho_t}}{p}{\mathbb{1}_{n^2}}$ cannot be expressed in terms of $\langle |E|^p \rangle$.  Last but not least, our basic integral form \QSL almost equals that of the operator-norm-based bound by Deffner in Ref.~\cite{Deffner17} for the case of $p = 2$ and $w = \mathbb{1}_{n^2}$ even though Deffner's result focused on infinite dimensional systems.  Here the difference is that Deffner expressed his \QSL via fidelity between the initial and final states while ours do not.  In addition, our bounds are completely different from his for other values of $p$ or $w$.

The Ky Fan's maximum principle (see Exercise~II.1.13 in Ref.~\cite{BhatiaMatrixAnalysis}) implies that
$s_1^\downarrow(A)^2 = \max_{\ket{j}} \bra{j} A A^\dag \ket{j} = \Symnormarrow{\!\VecOp(A A^\dag)}{p}{\mathbb{1}_1}$, where the maximum is over all normalized state $\ket{j}$.  So, maybe $\slbound[\text{int}]{p,\mathbb{1}_1,\mathcal{B}}$ and $\slbound[\text{sup}]{p,\mathbb{1}_1,\mathcal{B}}$ were representation basis independent.  However, these are not true because the states $\ket{j}$ that maximize $\Dnormarrowbasis{p}{w}{\mathcal{B}}(\rho_\tau,\rho_0)$ and $\Symnorm{\!\VecOp_\mathcal{B}(\mathbb{L} \rho_t)}{p}{\bar{w}}$ need not be the same.  In fact, our numerical study show that the only representation basis independent $\slbound[\text{int}]{p,w}$ and $\slbound[\text{sup}]{p,w}$ are the ones using $p = 2$ and $w = \mathbb{1}_{n^2}$.

	\subsubsection{Efficient Optimization over $w$}
	\label{Subsubsec:Properties_w}
	Interestingly, for a fixed $t$ and a fixed representation basis in computing $\rho_t$ together with $\rho_0$, the maximization of our \QSL{s} over $w$ requires only $n^2$ evaluations of the R.H.S. of Inequality~\eqref{E:new_QSL_group}.  We only need to consider the case of $p \in [1,\infty)$.  By taking the limit $p\to \infty$, our analysis also holds for the case of $p=\infty$.  Since our basic bounds in Inequality~\eqref{E:new_QSL_group} do not depend on the ordering of elements in $w$, without loss of generality, we only need to consider the case when elements of $w$ are non-negative and arranged in descending order.  Let $w = (w_i)$ and $w' = (w'_i)$ be two such sequences.  We study the behavior of $\slbound[\text{int}]{}$ and $\slbound[\text{sup}]{}$ when $w$ is replaced by $\lambda w + \bar{\lambda} w'$, where $\lambda \in [0,1]$ and $\bar{\lambda} = 1 - \lambda$.  The R.H.S. of both Eqs.~\eqref{E:new_QSL_integral} and~\eqref{E:new_QSL} can be written in the form $\displaystyle f(\lambda) = \left( \frac{\lambda N + \bar{\lambda} N'}{\lambda D + \bar{\lambda} D'} \right)^{1/p}$ with $D, D' > 0$ and $N, N' \ge 0$.  By elementary algebra, $f$ always attains its maximum value on the domain boundary of $\lambda$, namely $0$ or $1$.  That is to say, the maxima of Eqs.~\eqref{E:new_QSL_integral} and~\eqref{E:new_QSL} must be attained by using $w = \mathbb{1}_j$ for $j = 1,2,\cdots, n^2$. This is because only when $w$ is in this form that it cannot be decomposed as a weighted sum of two linearly independent vectors with non-negative elements arranged in descending order.  Consequently, to maximize our basic \QSL{s} for a fixed representation basis as well as $p$, we only need to evaluate the R.H.S. of Eqs.~\eqref{E:new_QSL_integral} and~\eqref{E:new_QSL} with $n^2$ different weights $w$.  This result is significant because, while the introduction of weightings improves the sharpness of the \QSL, it keeps the computational efficiency of our approach.

	\section{Applications} \label{Sec:Appl}

	The following five examples show the power and applicability of our \QSL{s} across a range of quantum systems.  Every curve in the figures below contains about $200$~data points.  Using the Manopt package in Matlab~\cite{JMLR:v15:boumal14a}, each such set of data points except for those concerning coherent state photon loss in Sec.~\ref{Subsec:Appl_Photon_Loss}, takes less than $10$~mins of runtime in a typical Laptop.  The coherent state photon loss simulation takes about $10$~hrs, most of time is used to search the global maximum solution.  Our Matlab codes are available through the link in Ref.~\cite{our_codes}.
	
	\subsection{Time-Independent Hamiltonian Evolution Of Pure States}
	\label{Subsec:Appl_Pure_State}
	We consider time-independent pure state evolution as warm up.
	\subsubsection{The Qubit Case}
	\label{Subsubsec:Appl_Qubit}
	Consider the evolution of the initial state $\ket{\psi_0} = (\ket{E_0} + \ket{E_1})/\sqrt{2}$ under the time-independent Hamiltonian
	\begin{equation}
	 H = \sum_j j \ket{E_j}\bra{E_j} ,
	 \label{E:time-indep_H_def}
	\end{equation}
	where $\{ \ket{E_j} \}$ is an orthonormal set of energy eigenstates.
	Clearly, $\Delta E$ of this system is $1/2$.  Moreover, it is well-known that the \MT bound $\slbound{\MT}$ is tight for this state when the evolution time $\tau$ lies in $[0,\pi\hbar]$.

	Fix a $\tau > 0$ and consider the basis $\mathcal{B}_d$ that diagonalizes $\Delta(\tau)$.  Then, $\rho_\tau = \frac{1}{2} \begin{pmatrix} 1 & e^{-i\tau/\hbar} \\ e^{i\tau/\hbar} & 1 \end{pmatrix}$, $\Delta_{\mathcal{B}_d}(\tau) = \Vecfun[\mathcal{B}_d]{\rho_\tau - \rho_0} = \begin{pmatrix} 0 & e^{-i\tau/\hbar}-1 & e^{i\tau/\hbar}-1 & 0 \end{pmatrix}^{\transpose}\!\!/2$.  As a result, $\Dnormarrowbasis{1}{\mathbb{1}_1}{\mathcal{B}_d}(\rho_\tau,\rho_0) = s^\downarrow_1(\Delta_{\mathcal{B}_d}(\tau)) = \max(0, |\frac{e^{-i\tau/\hbar}-1}{2}|, |\frac{e^{i\tau/\hbar}-1}{2}|, 0) = |\!\sin (\tau/2\hbar)|$ and the corresponding $\Symnorm{\!\VecOp_{\mathcal{B}_d}(\dot{\rho}(t))}{1}{\mathbb{\bar{1}}_1} = |\!\cos[(t-\tau/2)/\hbar]/2\hbar| \le 1/2\hbar$.  Hence, $\int_0^\tau \Symnorm{\!\VecOp_{\mathcal{B}_d}(\dot{\rho}(t))}{1}{\mathbb{\bar{1}}_1} dt = \Dnormarrowbasis{1}{\mathbb{1}_1}{\mathcal{B}_d}(\rho_\tau,\rho_0)$ whenever $\tau \in [0,\pi\hbar]$.  In other words, from Inequalities~\eqref{E:new_QSL_integral} and~\eqref{E:new_QSL}, $\slbound[\text{sup}]{1,\mathbb{1}_1,\mathcal{B}_d} = 2\hbar\sin(\tau/2\hbar)$ and $\slbound[\text{int}]{1,\mathbb{1}_1,\mathcal{B}_d} = \tau$ whenever $\tau \in [0,\pi\hbar]$.  So, our basic integral form \QSL is also tight for this system when $\tau \in [0,\pi\hbar]$.
In contrast, if we pick $\mathcal{B}$ to be the computational basis $\mathcal{B}_c$, by repeating the same calculation, we obtain $\Dnormarrowbasis{1}{\mathbb{1}_1}{\mathcal{B}_c}(\rho_\tau,\rho_0) = |\!\sin (\tau/2\hbar)|$ and $\Symnorm{\!\VecOp_{\mathcal{B}_c}(\dot{\rho}(t))}{1}{\mathbb{\bar{1}_1}} = 1/2\hbar$ for all $t$.  Therefore, both $\slbound[\text{sup}]{1,\mathbb{1}_1,\mathcal{B}_c}$ and $\slbound[\text{int}]{1,\mathbb{1}_1,\mathcal{B}_c}$ equal $2\hbar |\sin(\tau/2\hbar)|$, which are not tight whenever $\tau > 0$.  This shows the basis dependence of our \QSL{s}.

Surprisingly, in this simple example, for a randomly picked representation basis $\mathcal{B}$, there is at least a quarter chance that $\slbound[\text{int}]{1,\mathbb{1}_1,\mathcal{B}}$ is tight.  This is because $\mathcal{B}$ is obtained by applying the unitary transformation in the form
	 $U = \begin{pmatrix}
	  e^{i\phi} \cos \theta & e^{i\varphi} \sin\theta \\
	  -e^{-i\varphi} \sin\theta & e^{-i \phi} \cos\theta
	 \end{pmatrix}$ to $\mathcal{B}_d$ for some $\theta,\phi,\varphi \in \mathbb{R}$.
	 After some simple calculations, we obtain $\Dnormarrowbasis{1}{\mathbb{1}_1}{\mathcal{B}}(\rho_\tau,\rho_0) = |\!\cos 2\theta \sin(\tau/2)|$ if $|\!\cos 2\theta| \ge |\!\sin 2\theta|$ and $\Symnorm{\!\VecOp_\mathcal{B}(\dot{\rho}(t))}{1}{\mathbb{\bar{1}}_1} = |\!\cos 2\theta \cos[(t - \tau/2)/\hbar]  - \sin 2\theta \sin(\phi - \varphi) \sin [(t-\tau/2)/\hbar]| / 2\hbar$.  Thus, for randomly chosen $U$, there is at least a quarter chance that $\int_0^\tau \Symnorm{\!\VecOp_\mathcal{B}(\dot{\rho}(t))}{1}{\mathbb{\bar{1}}_1} dt = |\!\cos 2\theta| \sin(\tau/2)$ if $\tau \in [0,\pi\hbar]$.  And in this case, both the $|\!\cos 2\theta|$ factors in the numerator and denominator of $\slbound[\text{int}]{1,\mathbb{1}_1,\mathcal{B}}$ cancel each other, giving us the value of $\tau$ in the end.

	 We now increase $\tau$ to $4\pi\hbar/3 (> \pi\hbar)$.  We find numerically that $\tau > \slbound[\text{int}]{\text{opt}} = 3.20\hbar > \slbound{\MT} = 2\pi\hbar/3 = \slbound{\CZ}$.  Here, $\slbound{\CZ}$ is the optimized Chau-Zeng (\CZ) bound which is the best \QSL for closed system under time-independent evolution known to date~\cite{CZbound,CZbound_prog}.  Our optimized bound is attained when $w = \mathbb{1}_1$ (and hence for any $p \ge 1$).  Remarkably, just like the case of $\tau \in [0,\pi\hbar]$, numerous different representation bases can attain our optimized \QSL so that our optimized form \QSL can be found simply by Monte Carlo method with $\approx 100$~sampling bases.  We shall say more about this phenomenon in our upcoming applications.

	\subsubsection{The Four-Dimensional Qudit Case}
	\label{Subsubsec:Appl_Qudit}
	Since all known \QSL{s} for time-independent Hamiltonian in the literature can only be saturated in qubit or qutrit systems~\cite{CZbound}, we further show the power of our optimized form \QSL when applied to the following randomly chosen four-dimensional qudit system with no obvious symmetry or structure.  Let us evolve the initial state $\ket{\psi_0} = \sqrt{0.2} \ket{E_0} + \sqrt{0.4} \ket{E_1} + \sqrt{0.3} \ket{E_\pi} + \sqrt{0.1} \ket{E_4}$ under the $H$ in Eq.~\eqref{E:time-indep_H_def}.  At time $\tau$, the fidelity between the initial and final states equals $|0.2 + 0.4 e^{-i\tau/\hbar} + 0.3 e^{-i\pi\tau/\hbar} + 0.1 e^{-4i\tau/\hbar}|$.
Using this fidelity at $\tau = \slbound{\text{c}} = 3.43\hbar$ and by means of the publicly available program in Ref.~\cite{CZbound_prog}, we find that the optimized \CZ bound~\cite{CZbound} for this system is $0.87\hbar$.  And by setting $\mathcal{B}$ to be the energy eigenbasis, $\Symnorm{\!\VecOp(\mathbb{L} \rho_t)}{p}{\bar{w}}$ is time-independent for all $p$ and $\bar{w}$.  Moreover, our calculation shows that the best basic integral form \QSL $\slbound[\text{int}]{} = \slbound[\text{sup}]{} = 1.98\hbar$ occurs when $w = \mathbb{1}_1$ or ($\mathbb{1}_2$ and $p \ge 1$).  Obviously, this result is not impressive.

What surprises us is that $\slbound[\text{int}]{\text{opt}}$ agrees with $\tau$ up to at least 7~significant figures for any $\tau \in [0,\slbound{\text{c}}]$.  This strongly suggests that our \QSL $\slbound[\text{int}]{\text{opt}}$ is saturated for this system up to the evolution time $\slbound{\text{c}}$.  Since the \CZ bound is the most powerful bound for time-independent Hamiltonian evolution known up to this work~\cite{CZbound}, our pure state qubit and qudit time-independent Hamilton evolution examples demonstrate the power of our fully optimized form \QSL.
We find that generally the optimal sets of parameters appear to be degenerate and quite common in a lot of cases, especially when $\tau$ is small.  In fact, Monte Carlo method alone is sometimes sufficient to find $\slbound[\text{int}]{\text{opt}}$ or its approximation, lowering the actual computational complexity cost.  These observations are generic as they are found in all open and closed systems we have studied.  We further remark that for the qudit system we have studied here, it is possible to find $\tau > \slbound{\text{c}}$ such that $\slbound[\text{int}]{\text{opt}} = \tau$.  Moreover, the set of $\tau$'s for this to occur appears to be a fractal.  This deserves further investigations in future.

	\subsection{A Qubit Undergoing Spontaneous Emission}
	\label{Subsec:Appl_SpontEmission}
	
	Consider a qubit initialized to the state $(\ket{0}+\ket{1})/\sqrt{2}$.  We study its spontaneous emission to the ground state $\ket{0}$ under the dynamics governed by the Lindblad equation in the interaction picture~\cite{Deffner13}
	\begin{equation}
	 \frac{d\rho_t}{dt} = \mathbb{L}(\rho_t) = \gamma \left( \ket{0}\bra{1} \rho_t \ket{1}\bra{0} - \frac{1}{2} \{ \ket{1}\bra{1}, \rho_t \} \right) ,
		\label{eq:lindblad_spontaneous_emission}
	\end{equation}
	where $\gamma$ is the emission rate.  Surely, the density matrix of the qubit at time $t$ written in the energy eigenbasis is
	\begin{equation}
		\rho_t = \frac{1}{2} \begin{pmatrix} 2 - e^{-\gamma t} & e^{-\gamma t / 2} \\ e^{-\gamma t / 2} & e^{-\gamma t} \end{pmatrix}
		\label{eq:rho_t_spontaneous_emission}
	\end{equation}
	and the corresponding vectorized state equals $\Vecfun{\rho_t} = \begin{pmatrix} 2 - e^{-\gamma t} & e^{-\gamma t/2} & e^{-\gamma t/2} & e^{-\gamma t} \end{pmatrix}^{\transpose}\!\!/2$.
	
	 We compare the performance of our \QSL{s} with the one reported by Deffner and Lutz~\cite{Deffner13}, namely,
	\begin{equation}
	 \slbound{\DL} = \max\left( \frac{1}{\Lambda_{\tau}^{\text{op}}}, \frac{1}{\Lambda_{\tau}^{\text{tr}}}, \frac{1}{\Lambda_{\tau}^{\text{hs}}}\right) \cdot \sin^{2} \Bures(\rho_\tau, \rho_0) ,
	 \label{E:DL_bound}
	\end{equation}
	where $\Lambda_{\tau}^{\text{op/tr/hs}} = \frac{1}{\tau} \int_{0}^{\tau} \left\| \mathbb{L}(\rho_{t}) \right\|_{\text{op/tr/hs}} \ dt$ with op, tr and hr denotes the operator, trace and Hilbert-Schmidt norms, respectively.  As $\|\mathbb{L}(\rho_t)\|_{\text{tr}} = 2\|\mathbb{L}(\rho_t)\|_{\text{op}}$, Eq.~\eqref{E:DL_bound} equals $\slbound{\DL} = \max\left(1/\Lambda_{\tau}^{\text{op}}, 1/\Lambda_{\tau}^{\text{hs}}\right) \sin^2 \Bures(\rho_\tau,\rho_0)$.
	It turns out that the operator norm gives the tightest bound for this system~\cite{Deffner13}; and we refer to it as the \DL bound.  Moreover, the bound derived from the Hilbert-Schmidt norm serves as a generalization of the \MT bound for open systems.  Here we simply call it the \MT bound.

	As for our \QSL{s}, we first consider the case of $p = 1$, $w = \mathbb{1}_1$ and $\mathcal{B} = \mathcal{B}_E \equiv \{ \ket{0},\ket{1} \}$.  For any $\tau \ge 0$, it is clear from Eq.~\eqref{eq:rho_t_spontaneous_emission} that $\Dnormarrowbasis{1}{\mathbb{1}_1}{\mathcal{B}_E}(\rho_\tau,\rho_0) = (1 - e^{-\gamma \tau})/2$ and the corresponding $\Symnorm{\!\VecOp_{\mathcal{B}_E}(\dot{\rho_t})}{1}{\mathbb{\bar{1}}_1} = \gamma e^{-\gamma t}/2 \le \gamma/2$.  Hence, $\slbound[\text{sup}]{1,\mathbb{1}_1,\mathcal{B}_E} = (1 - e^{-\gamma \tau})/\gamma$ and $\slbound[\text{int}]{1,\mathbb{1}_1,\mathcal{B}_E} = \tau$.  In other words, our basic integral form \QSL is tight for all $\tau \ge 0$.  Actually, similar calculations give $\slbound[\text{sup}]{2,\mathbb{1}_1,\mathcal{B}_E} = \slbound[\text{sup}]{1,\mathbb{1}_1,\mathcal{B}_E}$ and $\slbound[\text{int}]{2,\mathbb{1}_1,\mathcal{B}_E} = \slbound[\text{int}]{1,\mathbb{1}_1,\mathcal{B}_E}$ so that $\slbound[\text{int}]{\text{opt}}$ may sometimes be attained using more than one set of parameters.

\begin{figure*}[t!]
	\centering
		\subfloat[\label{fig:qubit_opt}]{
			\includegraphics[width=0.47\textwidth]{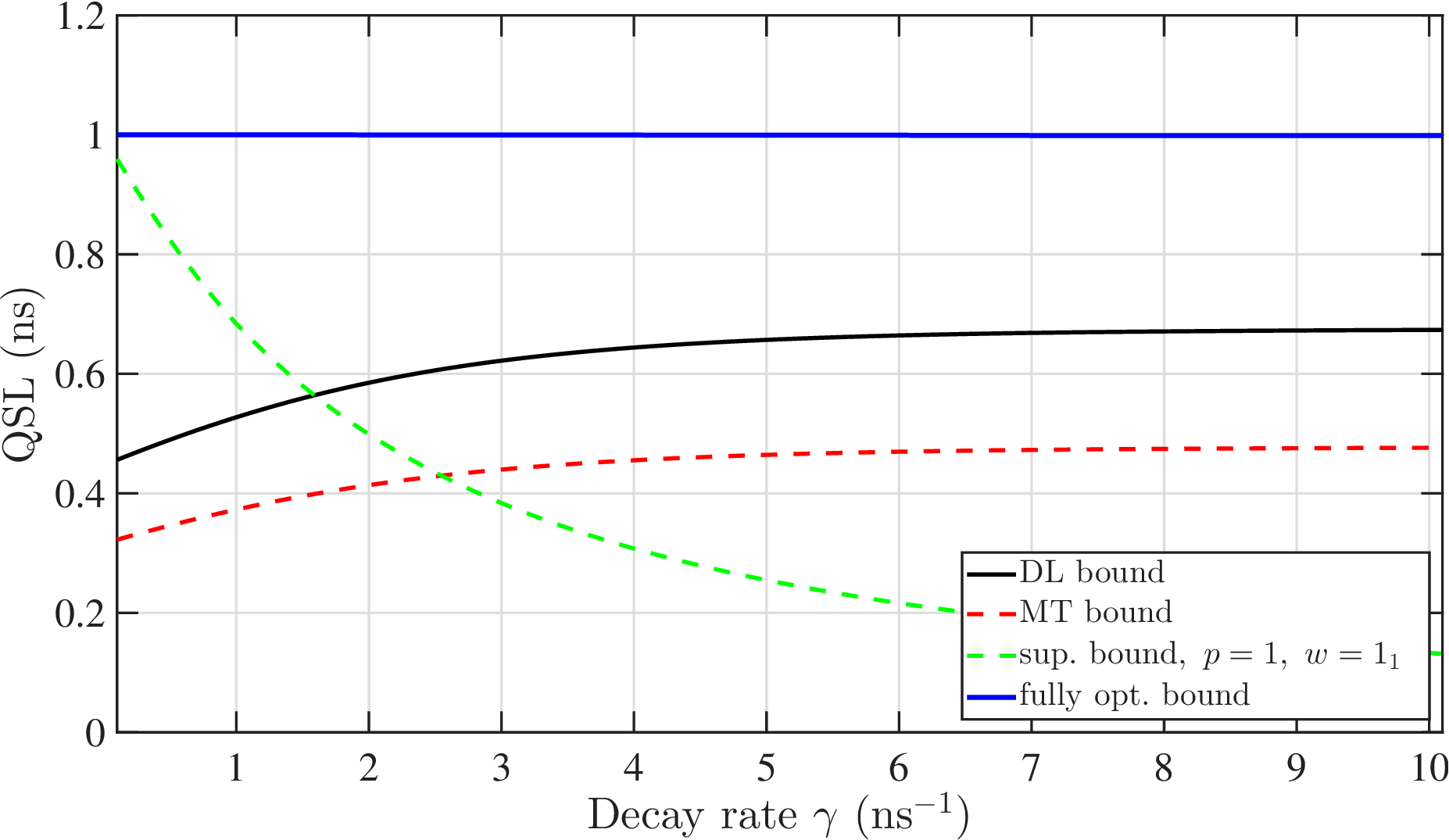}
		       }
	\hfill
		\subfloat[\label{fig:qubit_p2}]{
			\includegraphics[width=0.47\textwidth]{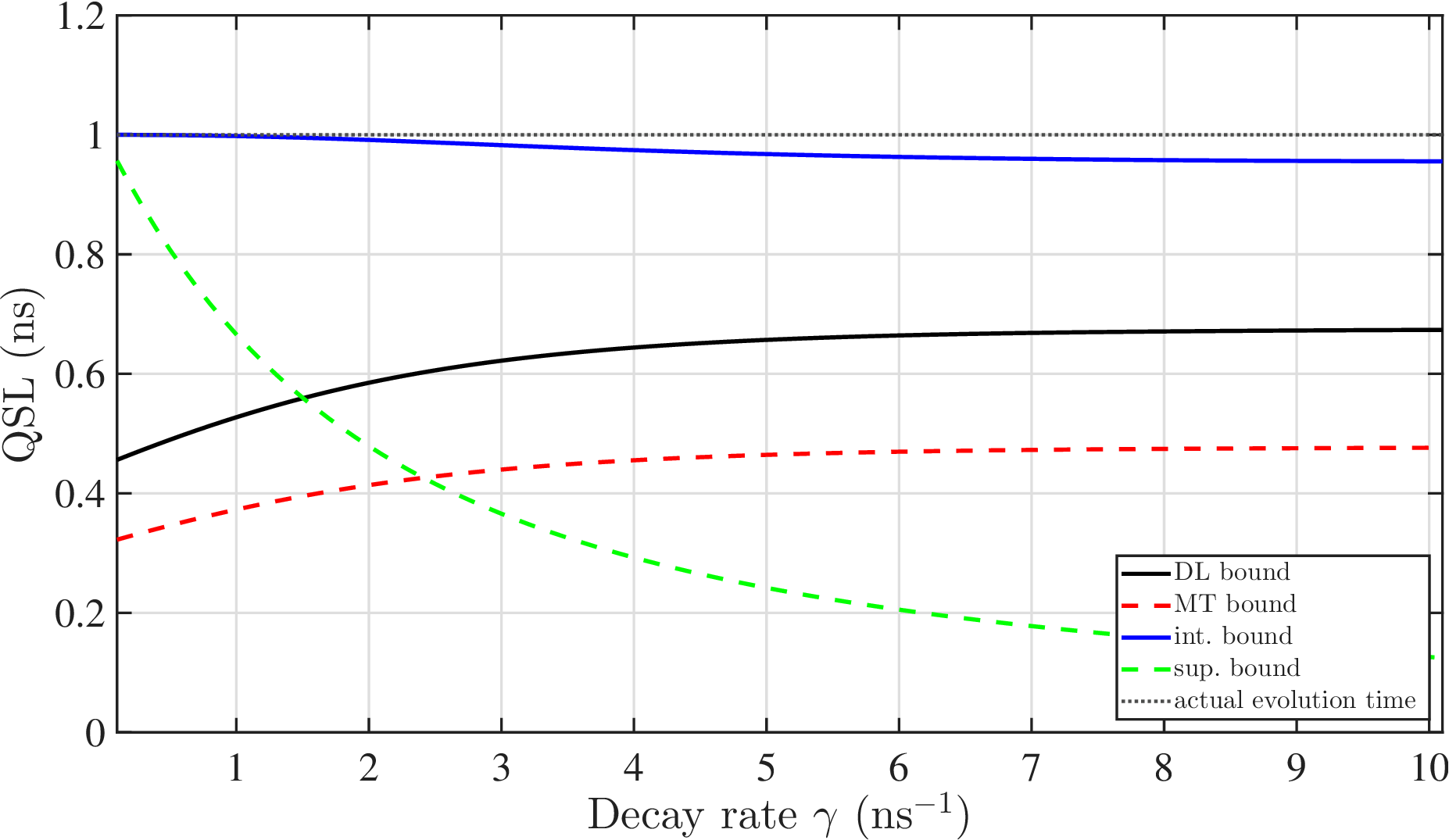}
		       }
        \\
		\subfloat[\label{fig:qubit_vary_p}]{
			\includegraphics[width=0.47\textwidth]{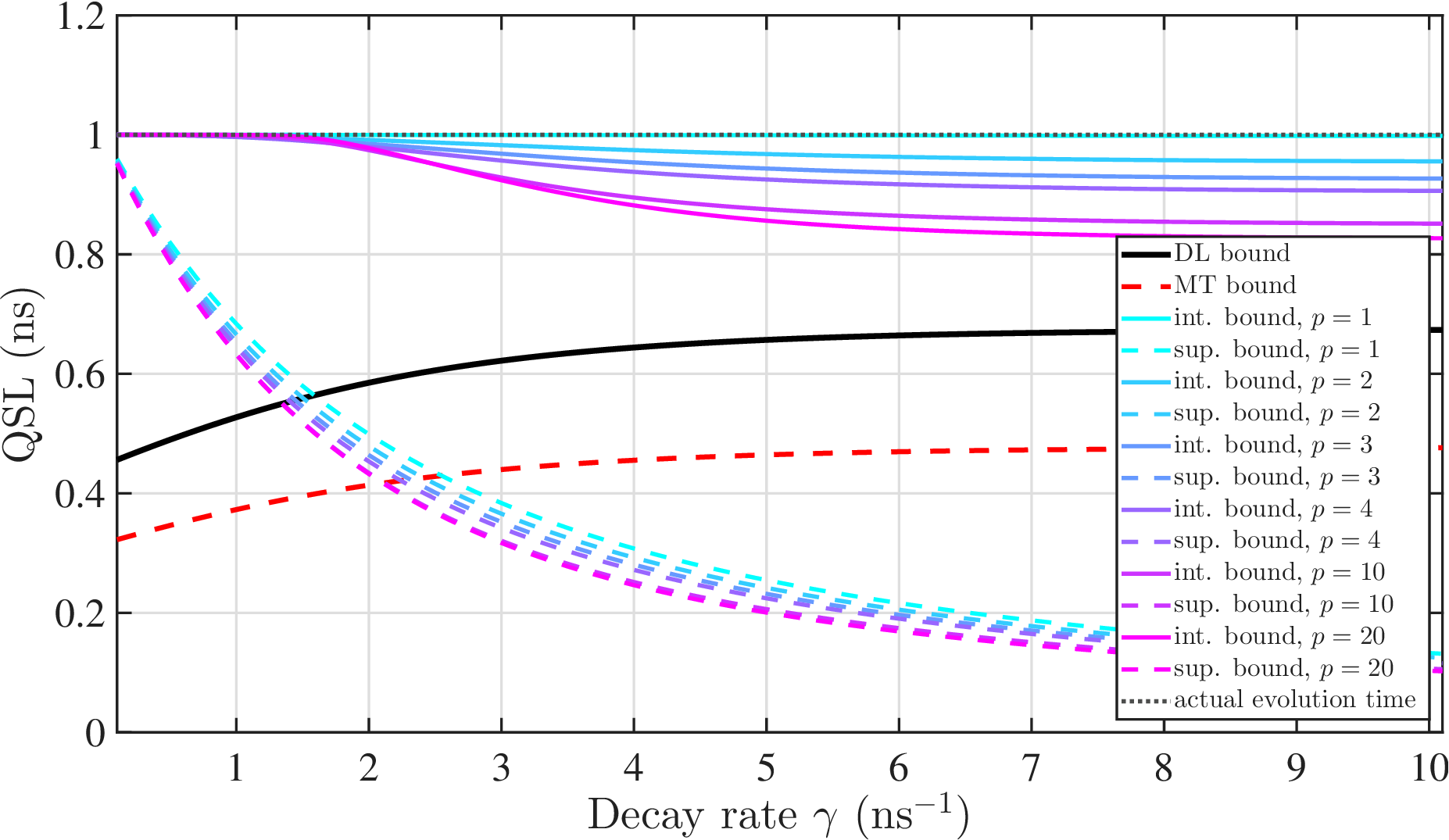}
		}
	\hfill
	\subfloat[\label{fig:qubit_vary_w}]{
			\includegraphics[width=0.47\textwidth]{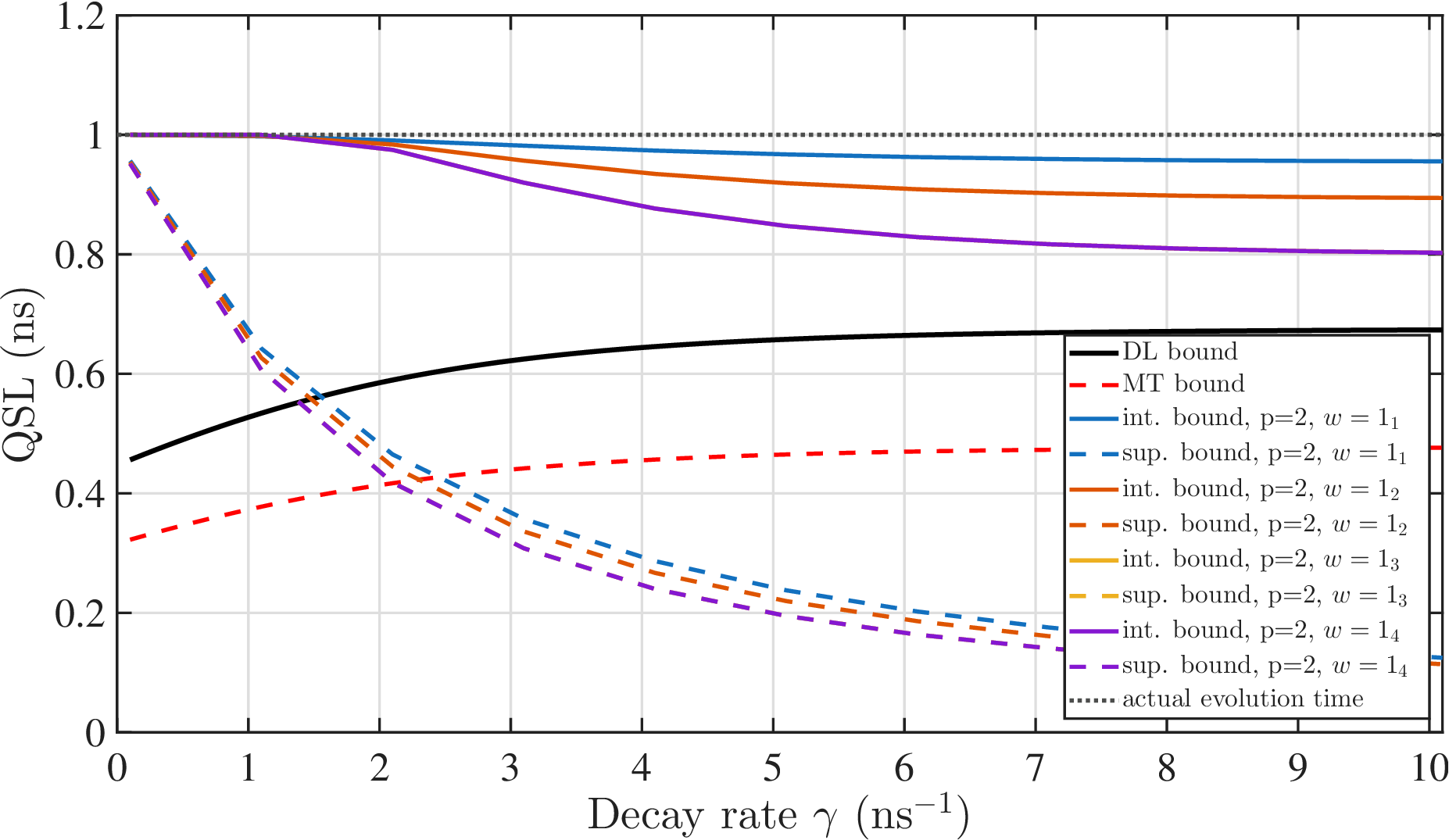}
		}
	\caption{Various \QSL{s} for a qubit undergoing spontaneous emission against the decay rate $\gamma$ when $\tau = 1$.  Panel~(a) compares $\slbound[\text{sup}]{1,\mathbb{1}_1,\mathcal{B}_E}$, $\slbound[\text{int}]{1,\mathbb{1}_1,\mathcal{B}_E}$ with the \MT and \DL bounds.  Note that $\slbound[\text{int}]{1,\mathbb{1}_1,\mathcal{B}_E}$ equals the actual evolution time $\tau$.
		Panel~(b) compares $\slbound[\text{sup}]{2,\mathbb{1}_4,\mathcal{B}_E}$ and $\slbound[\text{int}]{2,\mathbb{1}_4,\mathcal{B}_E}$ with the \MT and \DL bounds.
		Panels~(c) and~(d) vary $p$ and $w$ while keeping the other parameters fixed to those used in Panel~(b), respectively.
		\label{F:qubit_case}
	}
\end{figure*}

	To study the performance of our basic supremum form \QSL, we follow Ref.~\cite{Deffner13} by fixing the evolution time $\tau = 1$.  As shown in Fig.~\ref{fig:qubit_opt}, $\slbound[\text{sup}]{1,\mathbb{1}_1,\mathcal{B}_E}$ already beats both the \MT and \DL bounds when $\gamma < 1.6~\text{ns}^{-1}$ and $\gamma < 2.5~\text{ns}^{-1}$, respectively.  This shows the effectiveness of our supremum form \QSL as pointed out in Remark~\ref{Rem:sup}.

	To further investigate the dependence of our \QSL{s} on $p$, $w$ and $\mathcal{B}$, we consider the case of $\tau = 1$, $p = 2$, $w = \mathbb{1}_4$ and $\mathcal{B} = \mathcal{B}_E$.  As shown in Fig.~\ref{fig:qubit_p2}, the basic integral form \QSL $\slbound[\text{int}]{2,\mathbb{1}_4,\mathcal{B}_E}$ always wins whereas the basic supremum form \QSL $\slbound[\text{sup}]{2,\mathbb{1}_4,\mathcal{B}_E}$ beats the \DL and \MT bounds when $\gamma < 1.5~\text{ns}^{-1}$. and $\gamma < 2.4~\text{ns}^{-1}$, respectively.
Fig.~\ref{fig:qubit_vary_p} shows the effect of $p$ by fixing $p = 2$ and $\mathcal{B} = \mathcal{B}_E$ and Fig.~\ref{fig:qubit_vary_w} depicts the effect of varying $w$ while fixing the other two parameters.  Both figures highlights the effectiveness of our bounds over wide parameter ranges. 

To summarize, a significant advantage of our proposed bounds is that, even in the absence of full parameter optimization, the baseline formulation provides a remarkably tight constraint on the quantum evolution time. This is particularly valuable in scenarios where accurate system evolution is impossible and/or an exhaustive optimization search is either computationally intractable or experimentally unfeasible. Our results demonstrate that even for sub-optimal parameter choices, the bound remains close to the saturation (actual evolution time), affirming its practical utility.

	\subsection{High-fidelity Gates Using An \NV-center In Diamond}
	\label{Subsec:Appl_NV}
	
	Next, we consider the dynamics of spin-1 of a \NV center in diamond, a more complicated system that serves as a promising platform for quantum information processing. We analyze the time required to perform a high-fidelity quantum gate and compare our \QSL{s} with the \MT bound obtained in Ref.~\cite{Herb24}.
	
	\begin{figure*}[!t]
		\centering
		
		\subfloat[\label{fig:nv_opt}]{
			\includegraphics[width=\columnwidth]{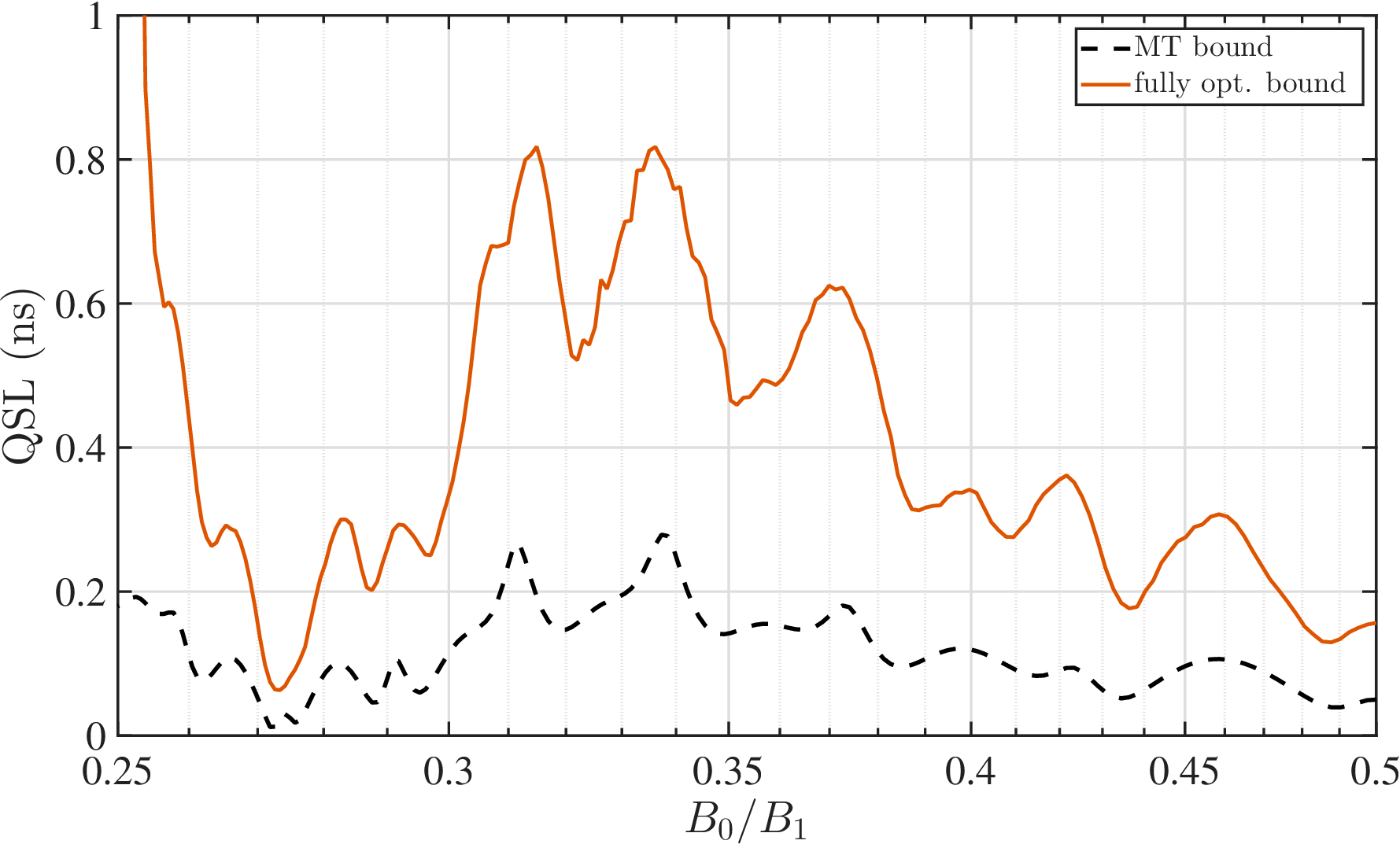}
		}
		\subfloat[\label{fig:nv_vary_p}]{
			\includegraphics[width=\columnwidth]{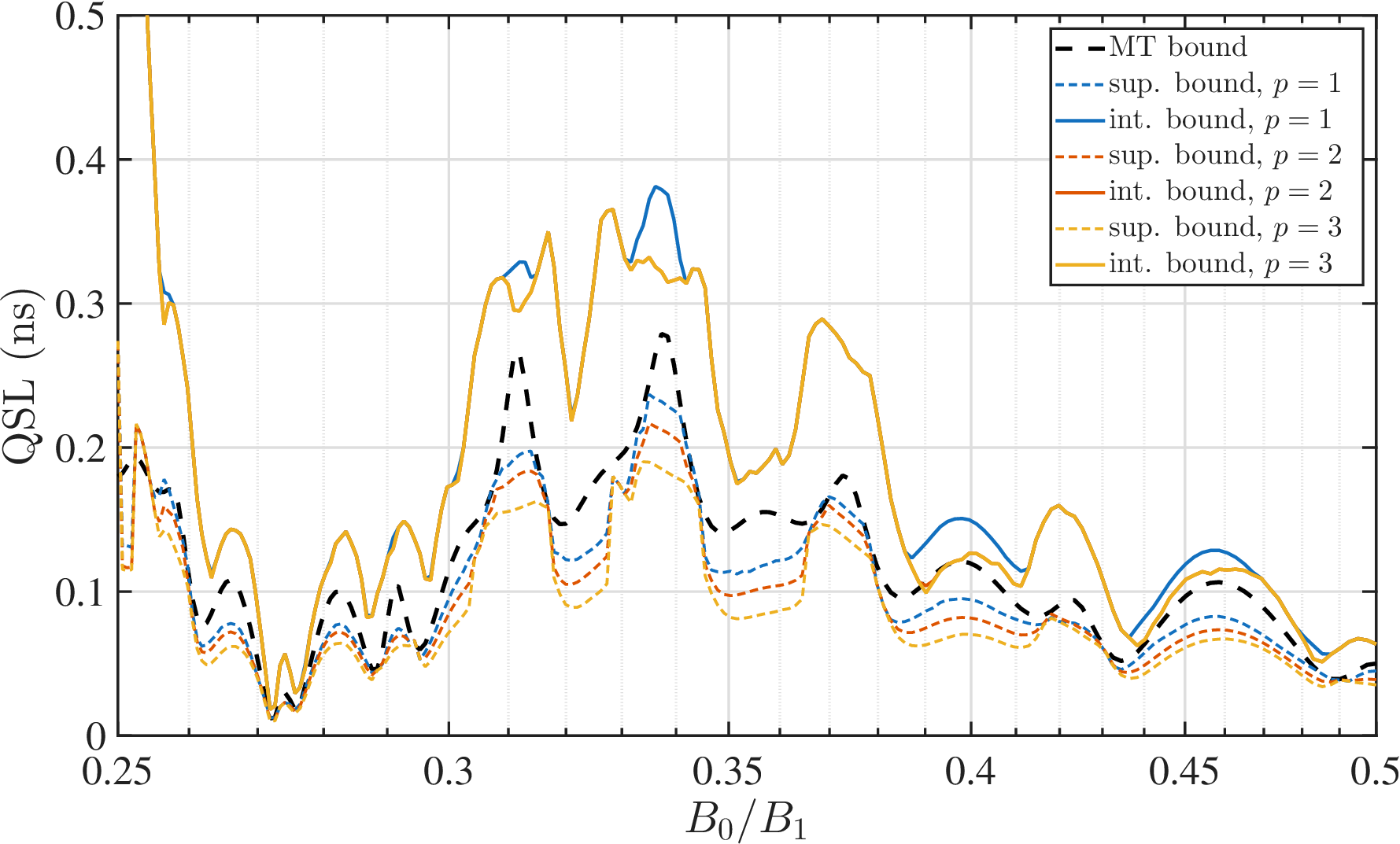}
		}
		\\
		\subfloat[\label{fig:nv_vary_w}]{
			\includegraphics[width=\columnwidth]{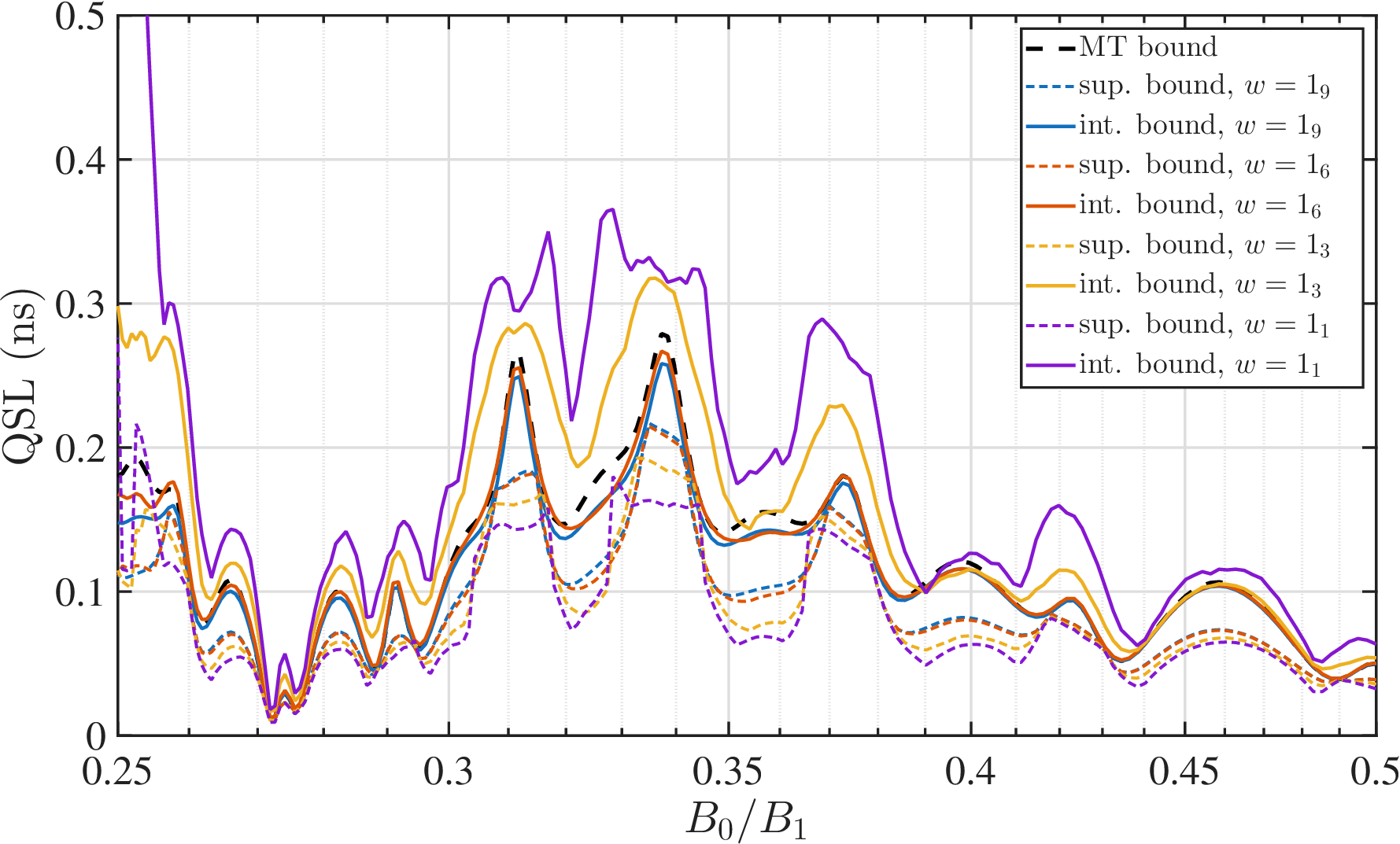}
		}
		\subfloat[\label{fig:nv_basis_opt}]{
			\includegraphics[width=\columnwidth]{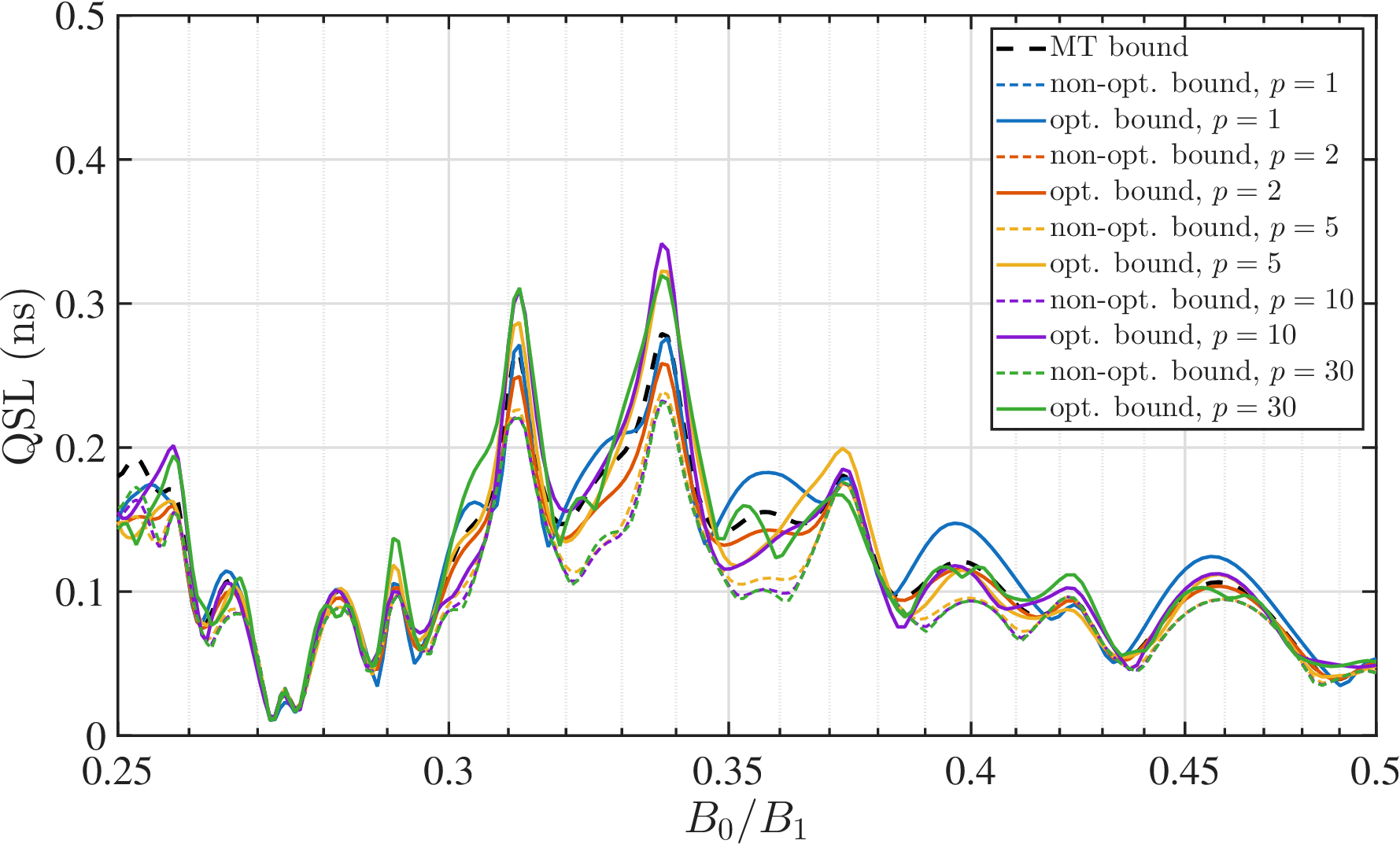}
		}
		
		\caption{Various \QSL{s} for the gate operation defined in Sec.~\ref{Subsec:Appl_NV} on an \NV-center spin against the magnetic field ratio $B_0/B_1$.  For simplicity, we set $\hbar = 1$ in all plots.
			Panel~(a) shows the fully optimized $\slbound[\text{int}]{\text{opt}}$ bound.  It uses a wider $y$-axis scale from the rest for better illustration.
			For Panel~(b), (c) and~(d), we vary the parameters $p$, $w$ and $\mathcal{B}$ by fixing the remaining parameters to $p=2$, $w=\mathbb{1}_9$ and $\mathcal{B}$ as the computational basis $\mathcal{B}_c$, respectively.
			\label{fig:qsl_nv_center_comparison}
		}
	\end{figure*}
	
	The spin operators $\hat{S}_x$, $\hat{S}_y$, and $\hat{S}_z$ in the basis $\{\ket{m_s = +1}, \ket{m_s = 0}, \ket{m_s = -1}\}$ are
	\begin{align}
		\hat{S}_x &= \frac{1}{\sqrt{2}}\begin{pmatrix} 0 & 1 & 0 \\ 1 & 0 & 1 \\ 0 & 1 & 0 \end{pmatrix}, \quad
		\hat{S}_y = \frac{1}{\sqrt{2}}\begin{pmatrix} 0 & -i & 0 \\ i & 0 & -i \\ 0 & i & 0 \end{pmatrix} \quad \text{and} \notag \\
		\hat{S}_z &= \begin{pmatrix} 1 & 0 & 0 \\ 0 & 0 & 0 \\ 0 & 0 & -1 \end{pmatrix}.
	\end{align}
	And the time-dependent Hamiltonian governing the spin dynamics is given by $\hat{H}(t) = \hat{H}_0 + \hat{H}_c(t)$, where $\hat{H}_0$ is the static Hamiltonian
	\begin{equation}
		\hat{H}_0 = \hbar ( D \hat{S}_z^2 + \gamma_e B_0 \hat{S}_z )
	\end{equation}
	with $D$ being the zero-field splitting parameter and $\gamma_e$ is the electron gyromagnetic ratio.  Besides, the control Hamiltonian $\hat{H}_c(t)$ representing the interaction with a time-dependent magnetic field component $B_1$ applied perpendicular to the \NV axis is given by
	\begin{equation}
		\hat{H}_c(t) = \hbar \gamma_e B_1 [ f_x(t) \hat{S}_x + f_y(t) \hat{S}_y ] ,
	\end{equation}
	with the control field direction switching at half the total evolution time $\tau$
	\begin{equation}
		(f_x(t), f_y(t)) = \begin{cases} (1,0) & \text{if } t < \tau/2 , \\ (0,1) & \text{if } t \ge \tau/2 . \end{cases}
	\end{equation}
	The system is initialized in the state $\ket{\psi_0} = \ket{m_s = +1}$. Here we set $D = 2\pi \times 2.87$~GHz, $\gamma_e = 2\pi \times 28.0345$~GHz/T and $B_0 = 0.05$~T in our computation.
	
	Fig.~\ref{fig:nv_opt} shows that $\slbound[\text{int}]{\text{opt}}$ wins the \MT bound for all values of $B_0/B_1$, often by a wide margin.  Moreover, the fully optimized $p$, $w$ and $\mathcal{B}$ depend on $B_0/B_1$ in a complicated way without obvious patterns or trends.
	Figs.~\ref{fig:nv_vary_p}, \ref{fig:nv_vary_w} and~\ref{fig:nv_basis_opt} show the dependence of $p$, $w$ and $\mathcal{B}$ by fixing the other parameters to $p = 2$, $w = \mathbb{1}_9$ and $\mathcal{B}$ equals the computational basis $\mathcal{B}_c$, respectively.  They tell us that optimizing either $p$, $w$ and $\mathcal{B}$ alone is already quite effective in getting a good bound.
	Further note that all bounds exhibit rather huge fluctuations over the magnetic field ratio $B_0/B_1$.  Nevertheless, our \QSL{s}, particularly for larger $p$, tend to show smaller variance. A small variance \QSL is experimentally advantageous, as it provides a more robust estimate of the minimal time required for high-fidelity gate operations, especially in the presence of minor parameter drifts~\cite{Herb24}. This analysis demonstrates that by optimizing over both $p$ and $w$, our \QSL{s} can offer a sharpness and less fluctuating bound than the existing method for complex, time-dependent quantum systems.
Finally, we remark that while our supremum form bounds are generally less effective than the \MT bound, they are close and very occasionally beats it.
	
\subsection{Photon Loss In An Infinite-Dimensional Bosonic Mode}
\label{Subsec:Appl_Photon_Loss}

\begin{figure*}[t]
	\centering
		\subfloat[\label{fig:photon_loss_p2}]{
			\includegraphics[width=0.47\textwidth]{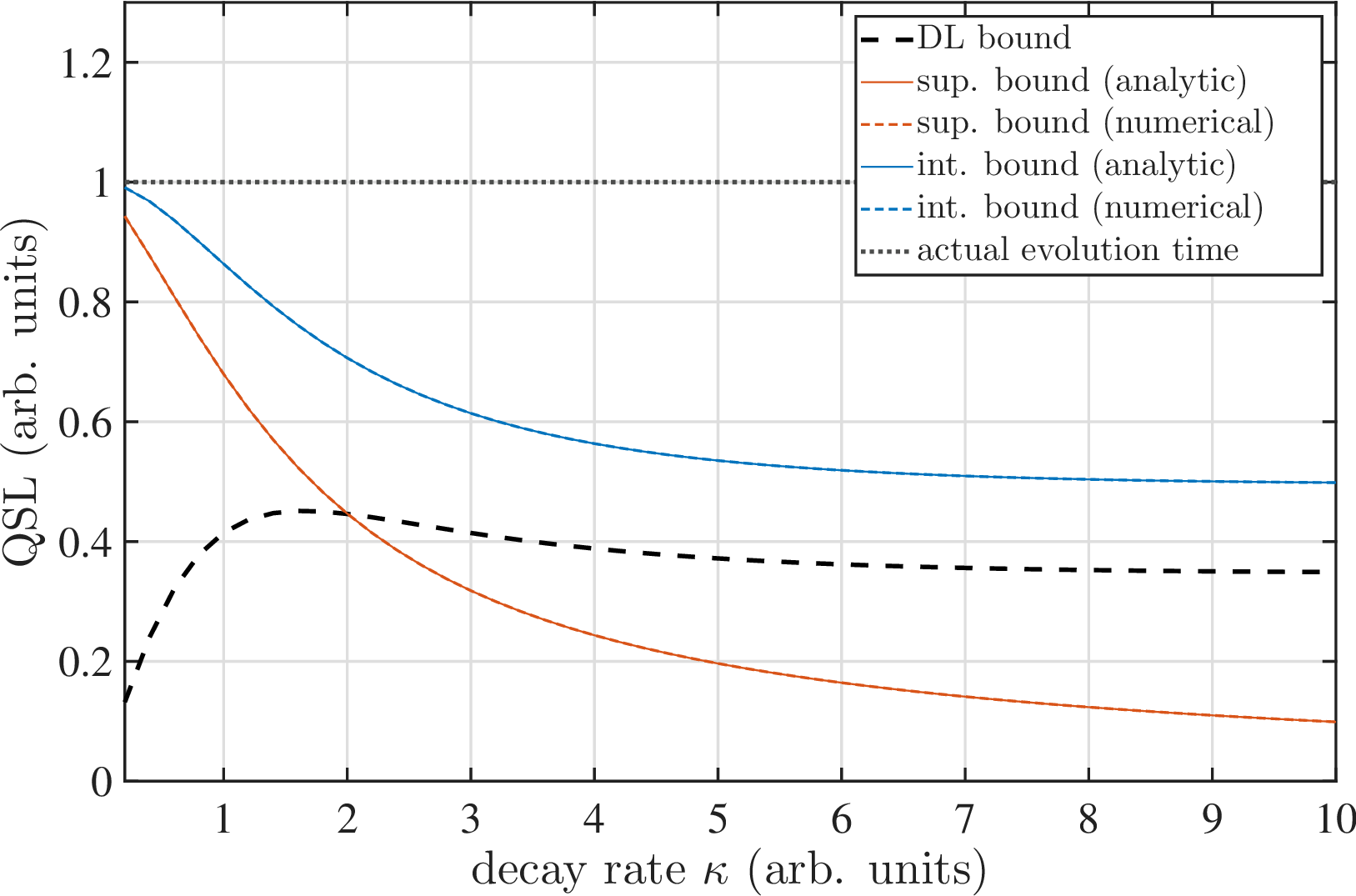}
		       }
	\hfill
		\subfloat[\label{fig:photon_loss_p3}]{
			\includegraphics[width=0.47\textwidth]{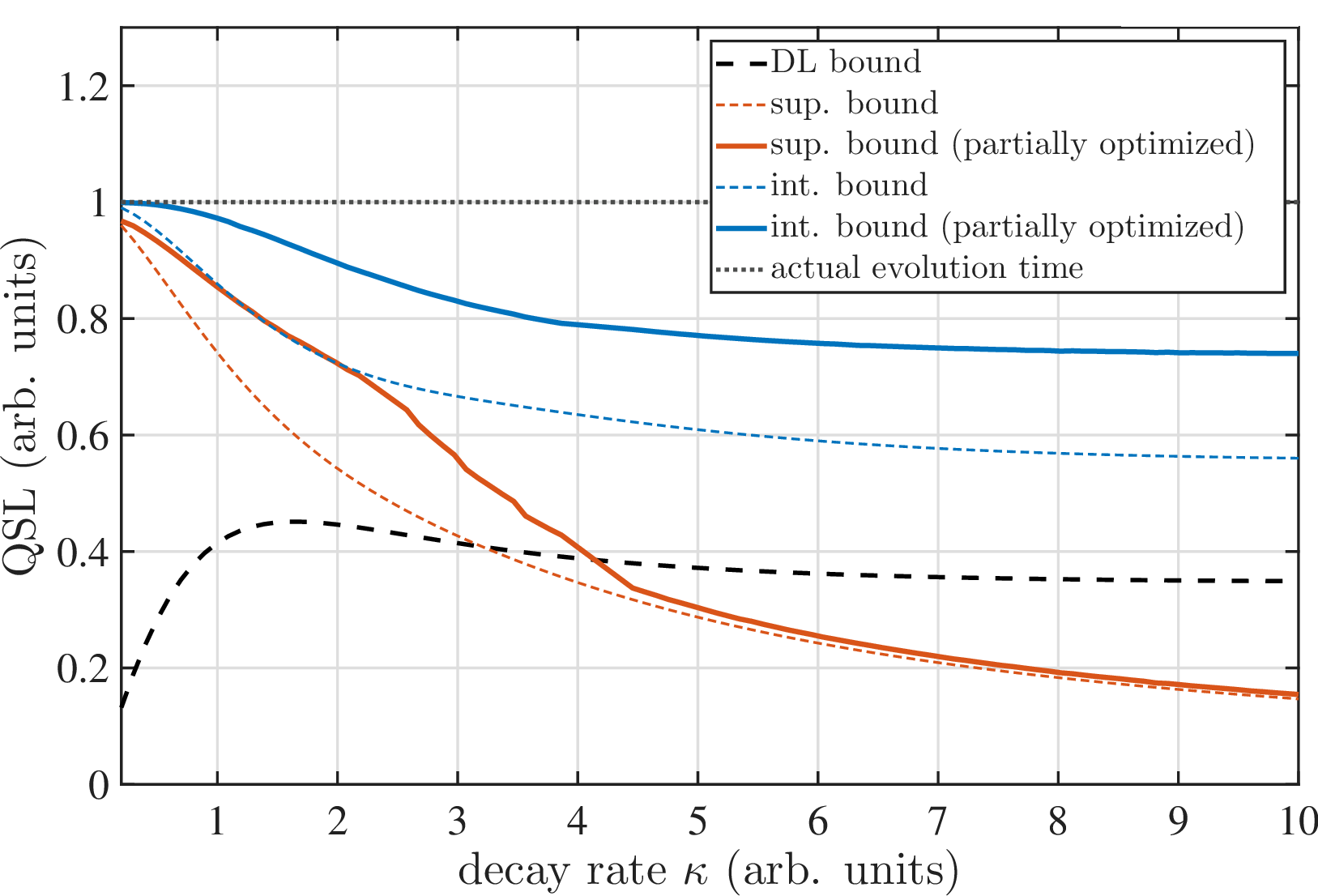}
		       }
	\caption{Various \QSL{s} for a bosonic mode undergoing photon loss plotted against the decay rate $\kappa$, with fixed evolution time $\tau=1$ and initial amplitude $\alpha_0=2$.  In Fig.~\ref{fig:photon_loss_p2}, which uses $p = 2$ and $w = \mathbb{1}$, thin solid lines are the analytical results while markers are numerical ones computed using the truncated Fock orthonormal set $\{ \ket{n} \}_{n=0}^{19}$.  In Fig.~\ref{fig:photon_loss_p3}, which uses $p = 3$ and $w = \mathbb{1}$, all results are numerically computed in the same truncated Fock orthonormal set.  The thin dashed and thick solid lines are for the basic bounds and the partially optimized ones defined in the text, respectively.}
	\label{fig:photon_loss}
\end{figure*}

Here we demonstrate the extensibility of our framework to countably infinite dimensional systems by considering a single bosonic mode undergoing photon loss. This is a paradigmatic open quantum system model describing dissipation in optical or microwave cavities. The dynamics is governed by the Lindblad master equation~\cite{BP07}
\begin{equation}
	\frac{d\rho_t}{dt}
	=
	\mathbb{L}\rho_t
	=
	\kappa\!\left(
	a\rho_t a^\dagger
	-
	\frac{1}{2}\{a^\dagger a,\rho_t\}
	\right),
	\label{eq:bosonic_loss_lindblad}
\end{equation}
where $a$ is the annihilation operator and $\kappa$ is the photon decay rate. The Hilbert space is spanned by the Fock states $\mathcal{B}_F \equiv \{\ket{n}\}_{n=0}^\infty$.
We initialize the system to the coherent state $\rho_0 = \ket{\alpha_0}\!\bra{\alpha_0}$ for some fixed $\alpha_0 \in \mathbb{C}$. Under the dynamics of Eq.~\eqref{eq:bosonic_loss_lindblad}, the state remains pure and coherent, evolving as $\rho_t = \ket{\alpha(t)}\!\bra{\alpha(t)}$ with amplitude $\alpha(t) = \alpha_0 e^{-\kappa t/2}$.

Interestingly, our basic \QSL{s} for $p=2$, and $w=\mathbb{1}$, namely, the uniform weight with all elements in the vector $\mathbb{1}$ being $1$, can be calculated analytically.  Note that the analysis in Sec.~\ref{Subsubsec:Properties_p=2} is valid for infinite dimensional systems so that $\slbound[\text{int}]{2,\mathbb{1},\mathcal{B}}$ and $\slbound[\text{sup}]{2,\mathbb{1},\mathcal{B}}$ are basis independent.  In fact, analogous to Eq.~\eqref{E:symnorm_2_1}, we know that
\begin{align}
 & \|\!\Vecfun[]{\rho_\tau-\rho_0}\|_{2,\mathbb{1}} \notag \\
	={}& \sqrt{\sum_j s^\downarrow_j(\rho_\tau - \rho_0)} 
	= \sqrt{\Tr[(\rho_\tau - \rho_0) (\rho_\tau - \rho_0)^\dag]} \notag \\
	={}& \sqrt{2 - 2 |\!\braket{\alpha(0)|\alpha(\tau)}\!|^2} \notag \\
	={}&
	\sqrt{
		2
		-
		2\exp\!\left[
		-|\alpha_0|^2
		\left(1-e^{-\kappa\tau/2}\right)^2
		\right]
	} .
	\label{eq:bosonic_distance}
\end{align}
To calculate $\Symnorm{\!\Vecfun[]{\mathbb{L}\rho_t}}{2}{\mathbb{1}}$, we observe that $\ket{\dot{\alpha}(t)} = \kappa [|\alpha(t)|^2 - \alpha(t) a^\dag] \ket{\alpha(t)}/2$ and hence $\braket{\dot{\alpha}(t)|\alpha(t)} = 0$.  Therefore,
\begin{align}
 \Symnorm{\!\Vecfun[]{\mathbb{L}\rho_t}}{2}{\mathbb{1}}
	&= \sqrt{\Tr( \dot{\rho}_t \dot{\rho}_t^\dag)} = \sqrt{2 \braket{\dot{\alpha}(t)|\dot{\alpha}(t)}} \notag \\
	&= \frac{\kappa |\alpha(t)|}{\sqrt{2}} = \frac{\kappa |\alpha_0|\,e^{-\kappa t/2}}{\sqrt{2}} \le \frac{\kappa |\alpha_0|}{\sqrt{2}}
	\label{eq:bosonic_speed}
\end{align}
for all $t \ge 0$.

We now proceed via Theorem~\ref{Thrm:extension}.  By means of the facts that $\braket{m| \mathbb{L}\rho_t | n}$ is smooth and decays exponentially for large $m,n$, it is straightforward to check that conditions in Theorem~\ref{Thrm:extension} hold for any $p \ge 1$, $w=\mathbb{1}$ and any basis $\mathcal{B}$.   This result allows us to compute our \QSL{s} with ease.  In particular, we get our two analytical \QSL{s}
\begin{equation}
 \slbound[\text{int}]{2,\mathbb{1}} \equiv \slbound[\text{int}]{2,\mathbb{1},\mathcal{B}} = \frac{\tau \sqrt{1 - \exp\left[-|\alpha_0|^2 \left(1 - e^{-\kappa \tau/2}\right)^2\right]}}{|\alpha_0|\left(1 - e^{-\kappa \tau/2}\right)} 
\label{E:QSL_int_photon_loss}
\end{equation}
and
\begin{equation}
 \slbound[\text{sup}]{2,\mathbb{1}} \equiv \slbound[\text{sup}]{2,\mathbb{1},\mathcal{B}} = \frac{2 \sqrt{1 - \exp\left[-|\alpha_0|^2 \left(1 - e^{-\kappa \tau/2}\right)^2\right]}}{\kappa |\alpha_0|}
 \label{E:QSL_sup_photon_loss}
\end{equation}
for any basis $\mathcal{B}$.  Whereas for a general $p$ and $\mathcal{B}$, our two \QSL{s} for $w = \mathbb{1}$ can be numerically computed.

Fig.~\ref{fig:photon_loss} compares our bounds with the \DL bound~\cite{Deffner13} for different decay rates $\kappa$ by fixing $\tau = 1$ and $\alpha_0 = 2$.
The thin solid lines in Fig.~\ref{fig:photon_loss_p2} are the analytical \QSL{s} stated in Eqs.~\eqref{E:QSL_int_photon_loss} and~\eqref{E:QSL_sup_photon_loss}, while the circles and squares denote the numerical values obtained by computing the corresponding \QSL{s} in the truncated Fock orthonormal set $\{ \ket{n} \}_{n=0}^{19}$.  We expect our numerical computational results to be a good approximation of the analytical result because the eigenenergies of this system are discrete and the states concerned are well approximated in this truncated Hilbert space~\cite{Peres}.  Indeed, the excellent agreement between these two verifies that our discrete approximation is numerically stable and converges rapidly to the true infinite dimensional system.  Because of this, all the numerical \QSL{s} plotted in Fig.~\ref{fig:photon_loss} are computed in this truncated Fock orthonormal set.
Consistent with Eq.~\eqref{E:new_QSL_group}, the integral forms $\slbound[\text{int}]{2,\mathbb{1}}$ and $\slbound[\text{int}]{3,\mathbb{1},\mathcal{B}_F}$ are always better than the corresponding supremum forms $\slbound[\text{sup}]{2,\mathbb{1}}$ and $\slbound[\text{sup}]{3,\mathbb{1},\mathcal{B}_F}$.  Besides, $\slbound[\text{sup}]{3,\mathbb{1},\mathcal{B}_F} \ge \slbound[\text{sup}]{2,\mathbb{1}}$ and $\slbound[\text{int}]{3,\mathbb{1},\mathcal{B}_F} \ge \slbound[\text{int}]{2,\mathbb{1}}$.  More importantly, these two basic integral form \QSL bounds consistently beat the \DL bound over the entire range of decay rates investigated. This highlights the power of our framework.
Furthermore, the thick solid lines in Fig.~\ref{fig:photon_loss_p3} is the \QSL obtained by optimizing over the basis $\mathcal{B}$ after fixing $p = 3$ and $w = \mathbb{1}$.  They show that optimizing over the basis alone yields significantly tighter bounds at the expense of a much longer runtime due to the huge number of local maxima encountered in the optimalization process. This confirms that our \QSL{s} extend naturally to countably infinite dimensional open systems.
	
	\subsection{Quantumness Measures}
	\label{Subsec:Appl_Quantumness}
	We now apply our \QSL{s} to study the non-commutativity of operators. Non-commutativity, quantified by the off-diagonal elements of one observable within the eigenbasis of another, is a measure of quantumness and a fundamental driver of state evolution in quantum dynamics~\cite{Ray22,Shrimali25}.  Here we use dephasing and coherence generation processes to demonstrate the versatility of our framework.  Specifically, we adopt the models from Ref.~\cite{Shrimali25} which derived separate \QSL{s} for these two processes. In contrast, our unified \QSL is applicable to both.
	
\subsubsection{The Dephasing Process}
First, we consider the dephasing process where the observable is initialized to $\mathcal{A}_0 = \sigma_y$ and evolved under the Hamiltonian $H = \sigma_x$. The time-evolved observable is given by
\begin{equation}
	\mathcal{A}_t = e^{-\gamma t/2}
	\begin{bmatrix}
		-\sin 2t & i(\frac{\gamma}{4} \sin 2t - \cos 2t) \\
		-i(\frac{\gamma}{4} \sin 2t - \cos 2t) & \sin 2t
	\end{bmatrix},
\end{equation}
and the corresponding value of ``quantumness'' between the observables $\mathcal{A}_t$ and $\mathcal{A}_0$ equals~\cite{Shrimali25}
\begin{equation}
 Q(\mathcal{A}_0, \mathcal{A}_t) \equiv 2 \|[\mathcal{A}_0,\mathcal{A}_t] \|^2 = 16 e^{-\gamma t} \sin^2 2t.
\end{equation}
For this specific process, the specialized \QSL
\begin{equation}
 T \geq T_Q \equiv \frac{\sqrt{Q(\mathcal{A}_0, \mathcal{A}_T)}}{\sqrt{2 \langle \| [\mathcal{A}_0, \mathcal{L}^\dagger(\mathcal{A}_t)] \|_{\text{hs}} \rangle_T}}
\end{equation}
has been derived in Ref.~\cite{Shrimali25}, where $\langle \cdot \rangle_T$ denotes the average of its argument over the time interval $[0,T]$.

\subsubsection{The Coherence Generation Process}
Second, we analyze coherence generation in a system initialized in the state $\rho_0 = \ket{0}\bra{0}$, which evolves via the Hamiltonian $H = \sigma_x$. The dynamics of the density matrix follows
\begin{equation}
	\frac{d\rho_t}{dt} = -i[H, \rho_t] + \frac{\gamma}{2} (\sigma_z \rho_t \sigma_z - \rho_t).
\end{equation}
At time $t$, the state is
	\begin{equation}
		\rho_t = \frac{1}{2} + e^{-\gamma t/2} \begin{bmatrix}
			\frac{1}{2} \cos 2t + \frac{\gamma}{8} \sin 2t & \frac{i}{2} \sin 2t \\
			-\frac{i}{2} \sin 2t & -\left(\frac{1}{2} \cos 2t + \frac{\gamma}{8} \sin 2t\right)
		\end{bmatrix}.
	\end{equation}
The coherence, measured with respect to the observable $\mathcal{A} = \sigma_z$, is found to be
\begin{equation}
 C(\rho_t, \mathcal{A}) \equiv -\frac{1}{2} \sum_k \Tr \{ \sqrt{\rho_t},\ket{k}\bra{k}] \}^2  = e^{-\gamma t/2} |\sin 2t|
\end{equation}
where the sum is over any orthonormal basis $\{ \ket{k} \}$ of the system.  The specialized bound
\begin{equation}
 T \geq T_C \equiv \frac{\sqrt{2} |\sqrt{C(\rho_0, \mathcal{A})} - \sqrt{C(\rho_T, \mathcal{A})}|}{\langle \sqrt{\sum_k ||\partial_t \sqrt{\rho_t} |k\rangle \langle k||^2} \rangle_T}
\end{equation}
has been found in Ref.~\cite{Shrimali25}.

\begin{figure}[t]
	\centering
	\includegraphics[width=\columnwidth]{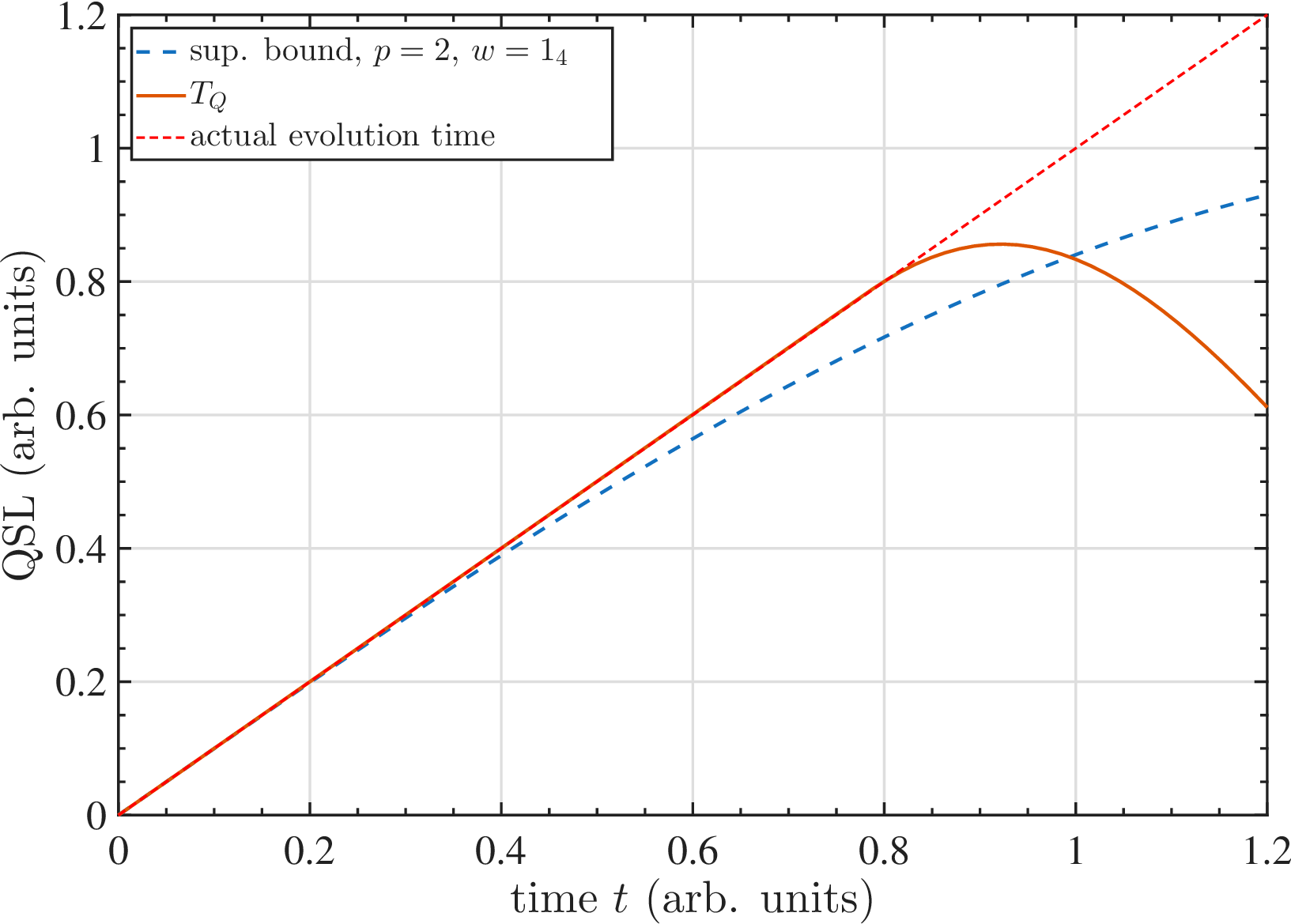}
	\caption{$\slbound[\text{sup}]{\text{opt}}$ and $T_Q$ for the qubit dephasing process.}
	\label{fig:qsl_dephasing}
\end{figure}

\begin{figure}[t]
	\centering
	\includegraphics[width=\columnwidth]{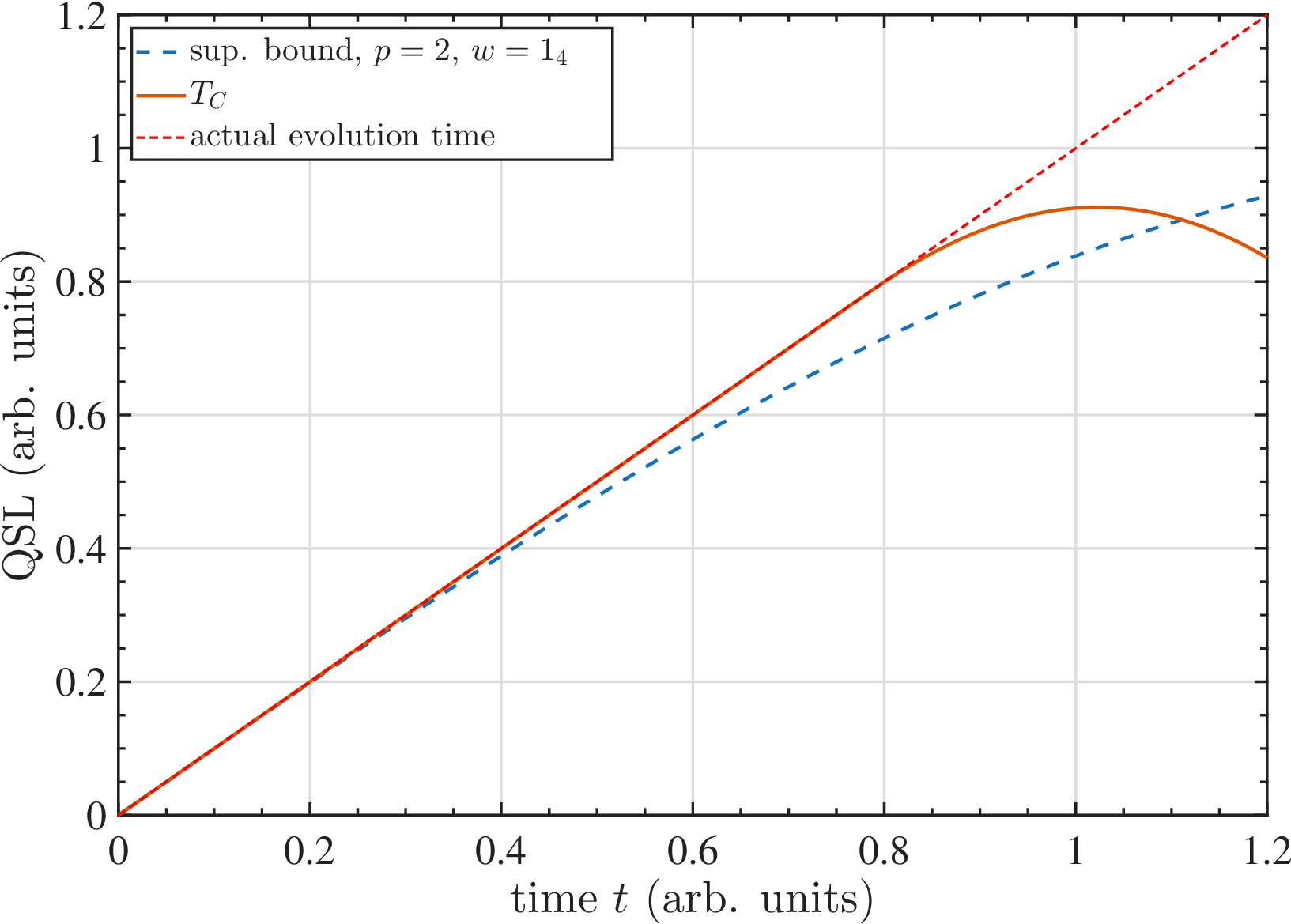}
	\caption{$\slbound[\text{sup}]{\text{opt}}$ and $T_C$ for the qubit coherence generation.}
	\label{fig:qsl_coherence_generation}
\end{figure}
	
We find that in both the dephasing and coherence generation processes, $\slbound[\text{int}]{\text{opt}}$ is attained at $p = 2$ and $w = \mathbb{1}_4$.  (Therefore, they are representation basis independent.)
	As shown in Figs.~\ref{fig:qsl_dephasing} and \ref{fig:qsl_coherence_generation}, our bounds are tight when $t < 0.35$ in both dephasing and coherence generation.  Moreover, they are better than the specialized bounds of Ref.~\cite{Shrimali25} for longer evolution times ($t > 1.0$ for dephasing and $t > 1.1$ for coherence generation). Although the specialized bounds are tighter for very short times, our \QSL{s} are more general, perform better at large $t$, do not decrease at large $t$, and more significantly, highlight the robustness of our framework in generalizing to operator dynamics beyond density matrices, a feature critical for addressing diverse quantum phenomena.
	
	\section{Discussions And Outlook}
	\label{Sec:Discussion}

	In summary, we have introduced a family of easy-to-use non-unitary-invariant $\Lnormarrow{p}{w}$-norm-based \QSL{s} that offers the benefits of universality and computational simplicity.
	Unlike most of the \QSL{s} on market, ours are basis representation dependent and are not monotone under the action of \CPTP maps.  Furthermore, just like a typical optimization problem, we do not find any obvious pattern on the choice of representation basis as well as the values of $p$ and $w$ in the optimized bound $\slbound[\text{int}]{\text{opt}}$.
	Demonstrating wide applicability, the family encompasses both closed and open quantum systems, under time-dependent and time-independent Hamiltonians. We have also extended the scope of the \QSL from density matrices to general quantum operators.
	As illustrated by the five provided examples, our family of \QSL{s} is powerful enough to give the best bounds known to date in many cases, particularly so when the evolution time $\tau$ is relatively short.
In fact, the optimized form $\slbound[\text{int}]{\text{opt}}$ is often tight when $\tau$ is small.  Besides, even the less model dependent and computationally less intensive basic supremum form $\slbound[\text{sup}]{}$ can be quite powerful in a lot of cases.
We believe that the use of the representation basis dependent $\Symnorm{\!\VecOp_\mathcal{B}(\mathbb{L} \rho_t)}{p}{\bar{w}}$ and its time integral are our keys of success.  In fact, using representation basis dependent norm is a unique feature among all the \QSL{s} that we aware of in the literature.

    From our five illustrative examples, we do not see any pattern on the optimized $p$, $w$ and $\mathcal{B}$ used.  Even for the case when $p$ and $w$ are fixed, we have no general analytical method to find the optimized $\mathcal{B}$.  Nevertheless, these numerical examples also show that representation basis optimization can be done in practice very quickly when $\tau$ is small because optimized basis are common.  Though we have studied the underlying reason analytically for the pure qubit state evolution under a time-independent Hamiltonian in Sec.~\ref{Subsubsec:Appl_Qubit}, a sound explanation for a general system is still lacking.  It is instructive to research on it in future.

\appendix
\section{Justification Of The Validity Of Eq.~\eqref{E:norm_extended_def}}
\label{Sec:norm_extension}
 We justify the validity of Eq.~\eqref{E:norm_extended_def} as follows.  First, let us consider the case when the R.H.S. of both Eqs.~\eqref{E:norm_extended_def1} and~\eqref{E:norm_extended_def2} are finite.  In this case, the series involved in both equations are absolutely convergent.  By rearranging the order of the sum, we may assume without loss of generality that $|\Delta_{\mathcal{B},j+1}(\tau)|^p = \left[ |\Delta_{\mathcal{B}}(\tau)|^\downarrow_{j+1} \right]^p$ for all $j\in \mathbb{N}$.  In this way, we see that there is a $P \in S_{\mathbb{N}}$ such that the sum in the R.H.S. of Eq.~\eqref{E:norm_extended_def1} equals that of the R.H.S. of Eq.~\eqref{E:norm_extended_def2}.  We now claim that if $P\in S_{\mathbb{N}}$ with $w_{P(j)+1} \ne w^\downarrow_{j+1}$ for some $j\in \mathbb{N}$, then $\displaystyle \sum_{j\in \mathbb{N}} w_{P(j)+1} |\Delta_{\mathcal{B},j+1}(\tau)|^p < \sum_{j\in \mathbb{N}} w^\downarrow_{j+1} \left[ |\Delta_{\mathcal{B}}(\tau)|^\downarrow_{j+1} \right]^p$.  Suppose $P$ is one such element in $S_{\mathbb{N}}$.  Let $\ell$ be the smallest element in $\mathbb{N}$ obeying $w_{P(\ell)+1} \ne w^\downarrow_{\ell+1}$ whose existence is guaranteed in our setting.  More importantly, $w_{P(\ell)+1} < w^\downarrow_{\ell+1}$.  Let $P' \in S_{\mathbb{N}}$ be the one sending $\ell$ to $P(\ell)$, $P(\ell)$ to $\ell$ and keeping all other elements fixed.  Obviously, $\displaystyle \sum_{j\in \mathbb{N}} w_{P(j)+1} |\Delta_{\mathcal{B},j+1}(\tau)|^p < \sum_{j\in \mathbb{N}} w_{P'(P(j))+1} |\Delta_{\mathcal{B},j+1}(\tau)|^p$.  This proves our claim as $P'\circ P \in S_{\mathbb{N}}$.  A direct consequence of this claim is that the maximum of Eq.~\eqref{E:norm_extended_def1} is attained when $w_{P(j)+1} = w^\downarrow_{j+1}$ for all $j\in \mathbb{N}$.  This proves the validity of Eq.~\eqref{E:norm_extended_def} if $\Dnormarrowbasis{p}{w}{\mathcal{B}}(\rho_\tau,\rho_0)$ is finite.  Using the above argument, it is easy to see that the R.H.S. of Eq.~\eqref{E:norm_extended_def1} is infinite if and only if R.H.S. of Eq.~\eqref{E:norm_extended_def2} is also infinite.  Therefore, the use of maximum in Eq.~\eqref{E:norm_extended_def1} is justified and Eq.~\eqref{E:norm_extended_def2} is valid in all cases.

 Finally, $\Dnormarrowbasis{p}{w}{\mathcal{B}}$ is clearly a semi-norm when $\rho_\tau$ and $\rho_0$ are countably infinite-dimensional density matrices.  It is a norm for whenever $\rho_\tau \ne \rho_0$, $|\Delta_{\mathcal{B}}(\tau)|^\downarrow_1 > 0$ and hence $\Dnormarrowbasis{p}{2}{\mathcal{B}}(\rho_\tau,\rho_0) > 0$.

\section{Proof Of Theorem~\ref{Thrm:extension}}
\label{Sec:proof_extension}
 Here we prove Theorem~\ref{Thrm:extension} by modifying the proof of Theorem~\ref{Thrm:main}.  By Condition~\ref{Cond:finiteness}, $\Dnormarrowbasis{p}{w}{\mathcal{B}}(\rho_\tau,\rho_0)$ is finite.  Hence, $\tau > 0$ and $\Dnormarrowbasis{p}{w}{\mathcal{B}}(\rho_t,\rho_0)$ and $\Dnormbasis{p}{\bar{w}}{\mathcal{B}}(\rho_t,\rho_0)$ can be regarded as functions of $t\in [0,\tau]$.  Next, using the notations in that proof, we need to show that for every $t\in J$, $\Dnormbasis{p}{\bar{w}}{\mathcal{B}}(\rho_t,\rho_0)$ is differentiable and
	 \begin{align}
	  \frac{d \Dnormarrowbasis{p}{w}{\mathcal{B}}}{dt} = \frac{d \Dnormbasis{p}{\bar{w}}{\mathcal{B}}}{dt} ={}& \Dnormbasis{p}{\bar{w}}{\mathcal{B}}^{1-p} \sum_{j\in \mathbb{N}} \bar{w}_{j+1} |\Delta_{\mathcal{B},j+1}|^{p-1} \times \notag \\
	  & \sgn(\Delta_{\mathcal{B},j+1}) \ \frac{d\!\Vecfun[\mathcal{B}]{\rho_t}_{j+1}}{dt} .
	  \label{E:dDdt_extended}
	\end{align}
 To do so, it suffices to prove that
 \begin{equation}
  \frac{d}{dt} \sum_{j\in {\mathbb N}} f_j(t) = \sum_{j\in \mathbb{N}} \frac{d f_j(t)}{dt}
  \label{E:diff_int_sign}
 \end{equation}
 whenever $t \in J$.  For any such $j$, clearly $df_j(t)/dt$ exists.  Moreover, Condition~\ref{Cond:finiteness} implies that $\sum_{j\in \mathbb{N}} f_j(t)$ converges for all $t\in [0,\tau]$.  Together with Condition~\ref{Cond:uniform_convergence} as well as Theorem~9.14 in Ref.~\cite{Apostol}, we conclude that Eq.~\eqref{E:diff_int_sign} and hence Eq.~\eqref{E:dDdt_extended} hold.

 Let $p > 1$ and $1/p + 1/q = 1$.  Applying Hölder inequality to the two vectors in Condition~\ref{Cond:Holder} and followed by multiplying both sides of the resultant inequality by $\Dnormbasis{p}{\bar{w}}{\mathcal{B}}^{1-p}$, we arrive at
	 \begin{equation}
	\left|\frac{d\Dnormarrowbasis{p}{w}{\mathcal{B}}}{dt}\right| \leq \left\|\frac{d\!\Vecfun[\mathcal{B}]{\rho_t}}{dt}\right\|_{p,\bar{w}}
	 = \Symnorm{\!\Vecfun[\mathcal{B}]{\mathbb{L} \rho_t}}{p}{\bar{w}}
	  \label{E:intermediate_extended}
	 \end{equation}
	 for all $t\in J$, where $\Symnorm{\cdot}{p}{\bar{w}}$ is the $\Lnorm{p}{\bar{w}}(\mathbb{N})$-seminorm.  Surely, Inequality~\eqref{E:intermediate_extended} holds for $p = 1$ or $\infty$ by taking the corresponding limits on $p$.

Finally, Condition~\ref{Cond:LDCT} and Lebesgue dominated convergence theorem imply the existence of $\displaystyle \int_0^\tau \Symnorm{\!\Vecfun[\mathcal{B}]{\mathbb{L} \rho_t}}{p}{\bar{w}} \ dt$.  This completes our proof of Theorem~\ref{Thrm:extension}.

\acknowledgments
This work is supported by the RGC Grant 17303323 of the HKSAR Government.

\nocite{apsrev42Control}
\bibliographystyle{apsrev4-2}
\bibliography{qsl3.4.bib}

\begin{thebibliography}{39}%
\makeatletter
\providecommand \@ifxundefined [1]{%
 \@ifx{#1\undefined}
}%
\providecommand \@ifnum [1]{%
 \ifnum #1\expandafter \@firstoftwo
 \else \expandafter \@secondoftwo
 \fi
}%
\providecommand \@ifx [1]{%
 \ifx #1\expandafter \@firstoftwo
 \else \expandafter \@secondoftwo
 \fi
}%
\providecommand \natexlab [1]{#1}%
\providecommand \enquote  [1]{``#1''}%
\providecommand \bibnamefont  [1]{#1}%
\providecommand \bibfnamefont [1]{#1}%
\providecommand \citenamefont [1]{#1}%
\providecommand \href@noop [0]{\@secondoftwo}%
\providecommand \href [0]{\begingroup \@sanitize@url \@href}%
\providecommand \@href[1]{\@@startlink{#1}\@@href}%
\providecommand \@@href[1]{\endgroup#1\@@endlink}%
\providecommand \@sanitize@url [0]{\catcode `\\12\catcode `\$12\catcode
  `\&12\catcode `\#12\catcode `\^12\catcode `\_12\catcode `\%12\relax}%
\providecommand \@@startlink[1]{}%
\providecommand \@@endlink[0]{}%
\providecommand \url  [0]{\begingroup\@sanitize@url \@url }%
\providecommand \@url [1]{\endgroup\@href {#1}{\urlprefix }}%
\providecommand \urlprefix  [0]{URL }%
\providecommand \Eprint [0]{\href }%
\providecommand \doibase [0]{https://doi.org/}%
\providecommand \selectlanguage [0]{\@gobble}%
\providecommand \bibinfo  [0]{\@secondoftwo}%
\providecommand \bibfield  [0]{\@secondoftwo}%
\providecommand \translation [1]{[#1]}%
\providecommand \BibitemOpen [0]{}%
\providecommand \bibitemStop [0]{}%
\providecommand \bibitemNoStop [0]{.\EOS\space}%
\providecommand \EOS [0]{\spacefactor3000\relax}%
\providecommand \BibitemShut  [1]{\csname bibitem#1\endcsname}%
\let\auto@bib@innerbib\@empty
\bibitem [{\citenamefont {Mandelstam}\ and\ \citenamefont
  {Tamm}(1945)}]{Mandelstam45}%
  \BibitemOpen
  \bibfield  {author} {\bibinfo {author} {\bibfnamefont {L.}~\bibnamefont
  {Mandelstam}}\ and\ \bibinfo {author} {\bibfnamefont {I.}~\bibnamefont
  {Tamm}},\ }\bibfield  {title} {\bibinfo {title} {{The uncertainty relation
  between energy and time in non-relativistic quantum mechanics}},\ }\href@noop
  {} {\bibfield  {journal} {\bibinfo  {journal} {J. Phys. (USSR)}\ }\textbf
  {\bibinfo {volume} {9}},\ \bibinfo {pages} {249} (\bibinfo {year}
  {1945})}\BibitemShut {NoStop}%
\bibitem [{\citenamefont {Braunstein}\ \emph {et~al.}(1996)\citenamefont
  {Braunstein}, \citenamefont {Caves},\ and\ \citenamefont
  {Milburn}}]{Braunstein96}%
  \BibitemOpen
  \bibfield  {author} {\bibinfo {author} {\bibfnamefont {S.~L.}\ \bibnamefont
  {Braunstein}}, \bibinfo {author} {\bibfnamefont {C.~M.}\ \bibnamefont
  {Caves}},\ and\ \bibinfo {author} {\bibfnamefont {G.~J.}\ \bibnamefont
  {Milburn}},\ }\bibfield  {title} {\bibinfo {title} {{Generalized uncertainty
  relations: Theory, examples, and Lorentz invariance}},\ }\href
  {https://doi.org/10.1006/aphy.1996.0040} {\bibfield  {journal} {\bibinfo
  {journal} {Ann. Phys.}\ }\textbf {\bibinfo {volume} {247}},\ \bibinfo {pages}
  {135} (\bibinfo {year} {1996})}\BibitemShut {NoStop}%
\bibitem [{\citenamefont {Margolus}\ and\ \citenamefont
  {Levitin}(1998)}]{Margolus98}%
  \BibitemOpen
  \bibfield  {author} {\bibinfo {author} {\bibfnamefont {N.}~\bibnamefont
  {Margolus}}\ and\ \bibinfo {author} {\bibfnamefont {L.~B.}\ \bibnamefont
  {Levitin}},\ }\bibfield  {title} {\bibinfo {title} {The maximum speed of
  dynamical evolution},\ }\href {https://doi.org/10.1016/S0167-2789(98)00054-2}
  {\bibfield  {journal} {\bibinfo  {journal} {Physica D}\ }\textbf {\bibinfo
  {volume} {120}},\ \bibinfo {pages} {188} (\bibinfo {year}
  {1998})}\BibitemShut {NoStop}%
\bibitem [{\citenamefont {Deffner}\ and\ \citenamefont
  {Lutz}(2013)}]{Deffner13}%
  \BibitemOpen
  \bibfield  {author} {\bibinfo {author} {\bibfnamefont {S.}~\bibnamefont
  {Deffner}}\ and\ \bibinfo {author} {\bibfnamefont {E.}~\bibnamefont {Lutz}},\
  }\bibfield  {title} {\bibinfo {title} {{Quantum speed limit for non-Markovian
  dynamics}},\ }\href {https://doi.org/10.1103/PhysRevLett.111.010402}
  {\bibfield  {journal} {\bibinfo  {journal} {Phys. Rev. Lett.}\ }\textbf
  {\bibinfo {volume} {111}},\ \bibinfo {pages} {010402} (\bibinfo {year}
  {2013})}\BibitemShut {NoStop}%
\bibitem [{\citenamefont {Deffner}\ and\ \citenamefont
  {Campbell}(2017)}]{Deffner17a}%
  \BibitemOpen
  \bibfield  {author} {\bibinfo {author} {\bibfnamefont {S.}~\bibnamefont
  {Deffner}}\ and\ \bibinfo {author} {\bibfnamefont {S.}~\bibnamefont
  {Campbell}},\ }\bibfield  {title} {\bibinfo {title} {{Quantum speed limits:
  From Heisenberg’s uncertainty principle to optimal quantum control}},\
  }\href {https://doi.org/10.1088/1751-8121/aa86c6} {\bibfield  {journal}
  {\bibinfo  {journal} {J. Phys. A}\ }\textbf {\bibinfo {volume} {50}},\
  \bibinfo {pages} {453001} (\bibinfo {year} {2017})}\BibitemShut {NoStop}%
\bibitem [{\citenamefont {Pratapsi}\ \emph {et~al.}(2025)\citenamefont
  {Pratapsi}, \citenamefont {Deffner},\ and\ \citenamefont
  {Gherardini}}]{PDG25}%
  \BibitemOpen
  \bibfield  {author} {\bibinfo {author} {\bibfnamefont {S.~S.}\ \bibnamefont
  {Pratapsi}}, \bibinfo {author} {\bibfnamefont {S.}~\bibnamefont {Deffner}},\
  and\ \bibinfo {author} {\bibfnamefont {S.}~\bibnamefont {Gherardini}},\
  }\bibfield  {title} {\bibinfo {title} {{Quantum speed limit for
  Kirkwood-Dirac quasiprobabilities}},\ }\href@noop {} {\bibfield  {journal}
  {\bibinfo  {journal} {Quant. Sci. Technol.}\ }\textbf {\bibinfo {volume}
  {10}},\ \bibinfo {pages} {035019} (\bibinfo {year} {2025})}\BibitemShut
  {NoStop}%
\bibitem [{\citenamefont {Pfeifer}\ and\ \citenamefont
  {Fröhlich}(1995)}]{Pfeifer95}%
  \BibitemOpen
  \bibfield  {author} {\bibinfo {author} {\bibfnamefont {P.}~\bibnamefont
  {Pfeifer}}\ and\ \bibinfo {author} {\bibfnamefont {J.}~\bibnamefont
  {Fröhlich}},\ }\bibfield  {title} {\bibinfo {title} {Generalized time-energy
  uncertainty relations and bounds on lifetimes of resonances},\ }\href
  {https://doi.org/10.1103/RevModPhys.67.759} {\bibfield  {journal} {\bibinfo
  {journal} {Rev. Mod. Phys.}\ }\textbf {\bibinfo {volume} {67}},\ \bibinfo
  {pages} {759} (\bibinfo {year} {1995})}\BibitemShut {NoStop}%
\bibitem [{\citenamefont {Anandan}\ and\ \citenamefont
  {Aharonov}(1990)}]{Anandan90}%
  \BibitemOpen
  \bibfield  {author} {\bibinfo {author} {\bibfnamefont {J.}~\bibnamefont
  {Anandan}}\ and\ \bibinfo {author} {\bibfnamefont {Y.}~\bibnamefont
  {Aharonov}},\ }\bibfield  {title} {\bibinfo {title} {Geometry of quantum
  evolution},\ }\href {https://doi.org/10.1103/PhysRevLett.65.1697} {\bibfield
  {journal} {\bibinfo  {journal} {Phys. Rev. Lett.}\ }\textbf {\bibinfo
  {volume} {65}},\ \bibinfo {pages} {1697} (\bibinfo {year}
  {1990})}\BibitemShut {NoStop}%
\bibitem [{\citenamefont {Pires}\ \emph {et~al.}(2016)\citenamefont {Pires},
  \citenamefont {Cianciaruso}, \citenamefont {Céleri}, \citenamefont
  {Adesso},\ and\ \citenamefont {Soares-Pinto}}]{Pires16}%
  \BibitemOpen
  \bibfield  {author} {\bibinfo {author} {\bibfnamefont {D.~P.}\ \bibnamefont
  {Pires}}, \bibinfo {author} {\bibfnamefont {M.}~\bibnamefont {Cianciaruso}},
  \bibinfo {author} {\bibfnamefont {L.~C.}\ \bibnamefont {Céleri}}, \bibinfo
  {author} {\bibfnamefont {G.}~\bibnamefont {Adesso}},\ and\ \bibinfo {author}
  {\bibfnamefont {D.~O.}\ \bibnamefont {Soares-Pinto}},\ }\bibfield  {title}
  {\bibinfo {title} {Generalized geometric quantum speed limits},\ }\href
  {https://doi.org/10.1103/PhysRevX.6.021031} {\bibfield  {journal} {\bibinfo
  {journal} {Phys. Rev. X}\ }\textbf {\bibinfo {volume} {6}},\ \bibinfo {pages}
  {021031} (\bibinfo {year} {2016})}\BibitemShut {NoStop}%
\bibitem [{\citenamefont {Taddei}\ \emph {et~al.}(2013)\citenamefont {Taddei},
  \citenamefont {Escher}, \citenamefont {Davidovich},\ and\ \citenamefont
  {de~Matos~Filho}}]{Taddei13}%
  \BibitemOpen
  \bibfield  {author} {\bibinfo {author} {\bibfnamefont {M.~M.}\ \bibnamefont
  {Taddei}}, \bibinfo {author} {\bibfnamefont {B.~M.}\ \bibnamefont {Escher}},
  \bibinfo {author} {\bibfnamefont {L.}~\bibnamefont {Davidovich}},\ and\
  \bibinfo {author} {\bibfnamefont {R.~L.}\ \bibnamefont {de~Matos~Filho}},\
  }\bibfield  {title} {\bibinfo {title} {Quantum speed limit for physical
  processes},\ }\href {https://doi.org/10.1103/PhysRevLett.110.050402}
  {\bibfield  {journal} {\bibinfo  {journal} {Phys. Rev. Lett.}\ }\textbf
  {\bibinfo {volume} {110}},\ \bibinfo {pages} {050402} (\bibinfo {year}
  {2013})}\BibitemShut {NoStop}%
\bibitem [{\citenamefont {de~Sousa}\ and\ \citenamefont {Pires}(2025)}]{SP25}%
  \BibitemOpen
  \bibfield  {author} {\bibinfo {author} {\bibfnamefont {J.~F.}\ \bibnamefont
  {de~Sousa}}\ and\ \bibinfo {author} {\bibfnamefont {D.~P}\ \bibnamefont
  {Pires}},\ }\bibfield  {title} {\bibinfo {title} {Generalized entropic
  quantum speed limits},\ }\href@noop {} {\bibfield  {journal} {\bibinfo
  {journal} {Phys. Rev. A}\ }\textbf {\bibinfo {volume} {112}},\ \bibinfo
  {pages} {012203} (\bibinfo {year} {2025})}\BibitemShut {NoStop}%
\bibitem [{\citenamefont {Mohan}\ \emph {et~al.}(2022)\citenamefont {Mohan},
  \citenamefont {Das},\ and\ \citenamefont {Pati}}]{MDP22}%
  \BibitemOpen
  \bibfield  {author} {\bibinfo {author} {\bibfnamefont {B.}~\bibnamefont
  {Mohan}}, \bibinfo {author} {\bibfnamefont {S.}~\bibnamefont {Das}},\ and\
  \bibinfo {author} {\bibfnamefont {A.~K.}\ \bibnamefont {Pati}},\ }\bibfield
  {title} {\bibinfo {title} {Quantum speed limits for information and
  coherence},\ }\href@noop {} {\bibfield  {journal} {\bibinfo  {journal} {New
  J. Phys.}\ }\textbf {\bibinfo {volume} {24}},\ \bibinfo {pages} {065003}
  (\bibinfo {year} {2022})}\BibitemShut {NoStop}%
\bibitem [{\citenamefont {Borrás}\ \emph {et~al.}(2006)\citenamefont
  {Borrás}, \citenamefont {Casas}, \citenamefont {Plastino},\ and\
  \citenamefont {Plastino}}]{BCPP06}%
  \BibitemOpen
  \bibfield  {author} {\bibinfo {author} {\bibfnamefont {A.}~\bibnamefont
  {Borrás}}, \bibinfo {author} {\bibfnamefont {M.}~\bibnamefont {Casas}},
  \bibinfo {author} {\bibfnamefont {A.~R.}\ \bibnamefont {Plastino}},\ and\
  \bibinfo {author} {\bibfnamefont {A.}~\bibnamefont {Plastino}},\ }\bibfield
  {title} {\bibinfo {title} {Entanglement dynamics and the quantum speed
  limit},\ }\href {https://doi.org/10.1103/PhysRevA.74.022326} {\bibfield
  {journal} {\bibinfo  {journal} {Phys. Rev. A}\ }\textbf {\bibinfo {volume}
  {74}},\ \bibinfo {pages} {022326} (\bibinfo {year} {2006})}\BibitemShut
  {NoStop}%
\bibitem [{\citenamefont {Fiderer}\ and\ \citenamefont
  {Braun}(2018)}]{Fiderer18}%
  \BibitemOpen
  \bibfield  {author} {\bibinfo {author} {\bibfnamefont {L.~J.}\ \bibnamefont
  {Fiderer}}\ and\ \bibinfo {author} {\bibfnamefont {D.}~\bibnamefont
  {Braun}},\ }\bibfield  {title} {\bibinfo {title} {Quantum metrology with
  quantum-chaotic sensors},\ }\href
  {https://doi.org/10.1038/s41467-018-03623-z} {\bibfield  {journal} {\bibinfo
  {journal} {Nature Commun.}\ }\textbf {\bibinfo {volume} {8}},\ \bibinfo
  {pages} {1351} (\bibinfo {year} {2018})}\BibitemShut {NoStop}%
\bibitem [{\citenamefont {Maleki}\ \emph {et~al.}(2023)\citenamefont {Maleki},
  \citenamefont {Ahansaz},\ and\ \citenamefont {Maleki}}]{Maleki23}%
  \BibitemOpen
  \bibfield  {author} {\bibinfo {author} {\bibfnamefont {Yusef}\ \bibnamefont
  {Maleki}}, \bibinfo {author} {\bibfnamefont {Bahram}\ \bibnamefont
  {Ahansaz}},\ and\ \bibinfo {author} {\bibfnamefont {Alireza}\ \bibnamefont
  {Maleki}},\ }\bibfield  {title} {\bibinfo {title} {Speed limit of quantum
  metrology},\ }\href {https://doi.org/10.1038/s41598-023-39082-w} {\bibfield
  {journal} {\bibinfo  {journal} {Sci. Rep.}\ }\textbf {\bibinfo {volume}
  {13}},\ \bibinfo {pages} {12031} (\bibinfo {year} {2023})}\BibitemShut
  {NoStop}%
\bibitem [{\citenamefont {Ness}\ \emph {et~al.}(2021)\citenamefont {Ness},
  \citenamefont {Lam}, \citenamefont {Alt}, \citenamefont {Meschede},
  \citenamefont {Sagi},\ and\ \citenamefont {Alberti}}]{Ness21}%
  \BibitemOpen
  \bibfield  {author} {\bibinfo {author} {\bibfnamefont {G.}~\bibnamefont
  {Ness}}, \bibinfo {author} {\bibfnamefont {M.~R.}\ \bibnamefont {Lam}},
  \bibinfo {author} {\bibfnamefont {W.}~\bibnamefont {Alt}}, \bibinfo {author}
  {\bibfnamefont {D.}~\bibnamefont {Meschede}}, \bibinfo {author}
  {\bibfnamefont {Y.}~\bibnamefont {Sagi}},\ and\ \bibinfo {author}
  {\bibfnamefont {A.}~\bibnamefont {Alberti}},\ }\bibfield  {title} {\bibinfo
  {title} {Observing crossover between quantum speed limits},\ }\href
  {https://doi.org/10.1126/sciadv.abj9119} {\bibfield  {journal} {\bibinfo
  {journal} {Sci. Adv.}\ }\textbf {\bibinfo {volume} {7}},\ \bibinfo {pages}
  {eabj9119} (\bibinfo {year} {2021})}\BibitemShut {NoStop}%
\bibitem [{\citenamefont {Pires}\ \emph {et~al.}(2024)\citenamefont {Pires},
  \citenamefont {deAzevedo}, \citenamefont {Soares-Pinto}, \citenamefont
  {Brito},\ and\ \citenamefont {Filgueiras}}]{Pires24}%
  \BibitemOpen
  \bibfield  {author} {\bibinfo {author} {\bibfnamefont {D.~P.}\ \bibnamefont
  {Pires}}, \bibinfo {author} {\bibfnamefont {E.~R.}\ \bibnamefont
  {deAzevedo}}, \bibinfo {author} {\bibfnamefont {D.~O.}\ \bibnamefont
  {Soares-Pinto}}, \bibinfo {author} {\bibfnamefont {F.}~\bibnamefont
  {Brito}},\ and\ \bibinfo {author} {\bibfnamefont {J.~G.}\ \bibnamefont
  {Filgueiras}},\ }\bibfield  {title} {\bibinfo {title} {Experimental
  investigation of geometric quantum speed limits in an open quantum system},\
  }\href {https://doi.org/10.1038/s42005-024-01634-5} {\bibfield  {journal}
  {\bibinfo  {journal} {Commun. Phys.}\ }\textbf {\bibinfo {volume} {7}},\
  \bibinfo {pages} {142} (\bibinfo {year} {2024})}\BibitemShut {NoStop}%
\bibitem [{\citenamefont {Cimmarusti}\ \emph {et~al.}(2015)\citenamefont
  {Cimmarusti}, \citenamefont {Yan}, \citenamefont {Patterson}, \citenamefont
  {Corcos}, \citenamefont {Orozco},\ and\ \citenamefont
  {Deffner}}]{Cimmarusti15}%
  \BibitemOpen
  \bibfield  {author} {\bibinfo {author} {\bibfnamefont {A.~D.}\ \bibnamefont
  {Cimmarusti}}, \bibinfo {author} {\bibfnamefont {Z.}~\bibnamefont {Yan}},
  \bibinfo {author} {\bibfnamefont {B.~D.}\ \bibnamefont {Patterson}}, \bibinfo
  {author} {\bibfnamefont {L.~P.}\ \bibnamefont {Corcos}}, \bibinfo {author}
  {\bibfnamefont {L.~A.}\ \bibnamefont {Orozco}},\ and\ \bibinfo {author}
  {\bibfnamefont {S.}~\bibnamefont {Deffner}},\ }\bibfield  {title} {\bibinfo
  {title} {Environment-assisted speed-up of the field evolution in cavity
  quantum electrodynamics},\ }\href
  {https://doi.org/10.1103/PhysRevLett.114.233602} {\bibfield  {journal}
  {\bibinfo  {journal} {Phys. Rev. Lett.}\ }\textbf {\bibinfo {volume} {114}},\
  \bibinfo {pages} {233602} (\bibinfo {year} {2015})}\BibitemShut {NoStop}%
\bibitem [{\citenamefont {Caneva}\ \emph {et~al.}(2009)\citenamefont {Caneva},
  \citenamefont {Murphy}, \citenamefont {Calarco}, \citenamefont {Fazio},
  \citenamefont {Montangero}, \citenamefont {Giovannetti},\ and\ \citenamefont
  {Santoro}}]{Caneva09}%
  \BibitemOpen
  \bibfield  {author} {\bibinfo {author} {\bibfnamefont {T.}~\bibnamefont
  {Caneva}}, \bibinfo {author} {\bibfnamefont {M.}~\bibnamefont {Murphy}},
  \bibinfo {author} {\bibfnamefont {T.}~\bibnamefont {Calarco}}, \bibinfo
  {author} {\bibfnamefont {R.}~\bibnamefont {Fazio}}, \bibinfo {author}
  {\bibfnamefont {S.}~\bibnamefont {Montangero}}, \bibinfo {author}
  {\bibfnamefont {V.}~\bibnamefont {Giovannetti}},\ and\ \bibinfo {author}
  {\bibfnamefont {G.~E.}\ \bibnamefont {Santoro}},\ }\bibfield  {title}
  {\bibinfo {title} {{Optimal control at the quantum speed limit}},\ }\href
  {https://doi.org/10.1103/PhysRevLett.103.240501} {\bibfield  {journal}
  {\bibinfo  {journal} {Phys. Rev. Lett.}\ }\textbf {\bibinfo {volume} {103}},\
  \bibinfo {pages} {240501} (\bibinfo {year} {2009})}\BibitemShut {NoStop}%
\bibitem [{\citenamefont {Giovannetti}\ \emph {et~al.}(2004)\citenamefont
  {Giovannetti}, \citenamefont {Lloyd},\ and\ \citenamefont
  {Maccone}}]{Giovannetti04}%
  \BibitemOpen
  \bibfield  {author} {\bibinfo {author} {\bibfnamefont {V.}~\bibnamefont
  {Giovannetti}}, \bibinfo {author} {\bibfnamefont {S.}~\bibnamefont {Lloyd}},\
  and\ \bibinfo {author} {\bibfnamefont {L.}~\bibnamefont {Maccone}},\
  }\bibfield  {title} {\bibinfo {title} {{Quantum-enhanced measurements:
  Beating the standard quantum limit}},\ }\href
  {https://doi.org/10.1126/science.1104149} {\bibfield  {journal} {\bibinfo
  {journal} {Science}\ }\textbf {\bibinfo {volume} {306}},\ \bibinfo {pages}
  {1330} (\bibinfo {year} {2004})}\BibitemShut {NoStop}%
\bibitem [{\citenamefont {Deffner}(2017)}]{Deffner17}%
  \BibitemOpen
  \bibfield  {author} {\bibinfo {author} {\bibfnamefont {S.}~\bibnamefont
  {Deffner}},\ }\bibfield  {title} {\bibinfo {title} {{Geometric quantum speed
  limits: A case for Wigner phase space}},\ }\href
  {https://doi.org/10.1088/1367-2630/aa83dc} {\bibfield  {journal} {\bibinfo
  {journal} {New J. Phys.}\ }\textbf {\bibinfo {volume} {19}},\ \bibinfo
  {pages} {103018} (\bibinfo {year} {2017})}\BibitemShut {NoStop}%
\bibitem [{\citenamefont {Herb}\ and\ \citenamefont {Degen}(2024)}]{Herb24}%
  \BibitemOpen
  \bibfield  {author} {\bibinfo {author} {\bibfnamefont {K.}~\bibnamefont
  {Herb}}\ and\ \bibinfo {author} {\bibfnamefont {C.~L.}\ \bibnamefont
  {Degen}},\ }\bibfield  {title} {\bibinfo {title} {{Quantum speed limit in
  quantum sensing}},\ }\href {https://doi.org/10.1103/PhysRevLett.133.210802}
  {\bibfield  {journal} {\bibinfo  {journal} {Phys. Rev. Lett.}\ }\textbf
  {\bibinfo {volume} {133}},\ \bibinfo {pages} {210802} (\bibinfo {year}
  {2024})}\BibitemShut {NoStop}%
\bibitem [{\citenamefont {Shrimali}\ \emph {et~al.}(2025)\citenamefont
  {Shrimali}, \citenamefont {Bhowmick},\ and\ \citenamefont
  {Pati}}]{Shrimali25}%
  \BibitemOpen
  \bibfield  {author} {\bibinfo {author} {\bibfnamefont {Divyansh}\
  \bibnamefont {Shrimali}}, \bibinfo {author} {\bibfnamefont {Swapnil}\
  \bibnamefont {Bhowmick}},\ and\ \bibinfo {author} {\bibfnamefont
  {Arun~Kumar}\ \bibnamefont {Pati}},\ }\bibfield  {title} {\bibinfo {title}
  {{Quantum speed limit on the production of quantumness of observables}},\
  }\href {https://doi.org/10.1103/PhysRevA.111.022445} {\bibfield  {journal}
  {\bibinfo  {journal} {Phys. Rev. A}\ }\textbf {\bibinfo {volume} {111}},\
  \bibinfo {pages} {022445} (\bibinfo {year} {2025})}\BibitemShut {NoStop}%
\bibitem [{\citenamefont {Chau}(2013)}]{Chau13}%
  \BibitemOpen
  \bibfield  {author} {\bibinfo {author} {\bibfnamefont {H.~F.}\ \bibnamefont
  {Chau}},\ }\bibfield  {title} {\bibinfo {title} {Quantum speed limit with
  forbidden speed intervals},\ }\href@noop {} {\bibfield  {journal} {\bibinfo
  {journal} {Phys. Rev. A}\ }\textbf {\bibinfo {volume} {87}},\ \bibinfo
  {pages} {052142} (\bibinfo {year} {2013})}\BibitemShut {NoStop}%
\bibitem [{\citenamefont {Watrous}(2018)}]{Watrous}%
  \BibitemOpen
  \bibfield  {author} {\bibinfo {author} {\bibfnamefont {J.}~\bibnamefont
  {Watrous}},\ }\href@noop {} {\emph {\bibinfo {title} {The Theory Of Quantum
  Information}}}\ (\bibinfo  {publisher} {CUP},\ \bibinfo {address}
  {Cambridge},\ \bibinfo {year} {2018})\ \bibinfo {note} {{E}qs.~(1.127)
  and~(1.128)}\BibitemShut {NoStop}%
\bibitem [{\citenamefont {Bhatia}(1997)}]{BhatiaMatrixAnalysis}%
  \BibitemOpen
  \bibfield  {author} {\bibinfo {author} {\bibfnamefont {Rajendra}\
  \bibnamefont {Bhatia}},\ }\href {https://doi.org/10.1007/978-1-4612-0653-8_4}
  {\emph {\bibinfo {title} {{Matrix Analysis}}}}\ (\bibinfo  {publisher}
  {Springer-Verlag},\ \bibinfo {year} {1997})\BibitemShut {NoStop}%
\bibitem [{\citenamefont {Chau}(2010)}]{Chau10}%
  \BibitemOpen
  \bibfield  {author} {\bibinfo {author} {\bibfnamefont {H.~F.}\ \bibnamefont
  {Chau}},\ }\bibfield  {title} {\bibinfo {title} {{Tight upper bound of the
  maximum speed of evolution of a quantum state}},\ }\href@noop {} {\bibfield
  {journal} {\bibinfo  {journal} {Phys. Rev. A}\ }\textbf {\bibinfo {volume}
  {81}},\ \bibinfo {pages} {062133} (\bibinfo {year} {2010})}\BibitemShut
  {NoStop}%
\bibitem [{\citenamefont {Chau}\ and\ \citenamefont {Lee}(2013)}]{LeeChau13}%
  \BibitemOpen
  \bibfield  {author} {\bibinfo {author} {\bibfnamefont {H.~F.}\ \bibnamefont
  {Chau}}\ and\ \bibinfo {author} {\bibfnamefont {K.-Y.}\ \bibnamefont {Lee}},\
  }\bibfield  {title} {\bibinfo {title} {{Relation between quantum speed limits
  and metrics on $U(n)$}},\ }\href@noop {} {\bibfield  {journal} {\bibinfo
  {journal} {J. Phys. A}\ }\textbf {\bibinfo {volume} {46}},\ \bibinfo {pages}
  {015305} (\bibinfo {year} {2013})}\BibitemShut {NoStop}%
\bibitem [{\citenamefont {Smith}(1993)}]{Smith93}%
  \BibitemOpen
  \bibfield  {author} {\bibinfo {author} {\bibfnamefont {S.~T.}\ \bibnamefont
  {Smith}},\ }\emph {\bibinfo {title} {Geometric Optimization Methods For
  Adaptive Filtering}},\ \href@noop {} {Ph.D. thesis},\ \bibinfo  {school}
  {Harvard U.} (\bibinfo {year} {1993})\BibitemShut {NoStop}%
\bibitem [{\citenamefont {Taylor}\ and\ \citenamefont {Kriegman}(1994)}]{TK94}%
  \BibitemOpen
  \bibfield  {author} {\bibinfo {author} {\bibfnamefont {C.~J.}\ \bibnamefont
  {Taylor}}\ and\ \bibinfo {author} {\bibfnamefont {D.~J.}\ \bibnamefont
  {Kriegman}},\ }\href@noop {} {\emph {\bibinfo {title} {{Minimization of the
  Lie group $SO(3)$ and related manifolds}}}},\ \bibinfo {type} {Tech. Rep.}\
  \bibinfo {number} {No.~9405}\ (\bibinfo  {institution} {Yale University},\
  \bibinfo {year} {1994})\BibitemShut {NoStop}%
\bibitem [{\citenamefont {Tao}\ and\ \citenamefont {Ohsawa}(2020)}]{TO20}%
  \BibitemOpen
  \bibfield  {author} {\bibinfo {author} {\bibfnamefont {M.}~\bibnamefont
  {Tao}}\ and\ \bibinfo {author} {\bibfnamefont {T.}~\bibnamefont {Ohsawa}},\
  }\bibfield  {title} {\bibinfo {title} {{Variational optimization on Lie
  group, with examples of leading (generalized) eigenvalue problems}},\
  }\href@noop {} {\bibfield  {journal} {\bibinfo  {journal} {Proc. Mach. Learn.
  Res.}\ }\textbf {\bibinfo {volume} {108}},\ \bibinfo {pages} {4269} (\bibinfo
  {year} {2020})}\BibitemShut {NoStop}%
\bibitem [{\citenamefont {Boumal}\ \emph {et~al.}(2014)\citenamefont {Boumal},
  \citenamefont {Mishra}, \citenamefont {Absil},\ and\ \citenamefont
  {Sepulchre}}]{JMLR:v15:boumal14a}%
  \BibitemOpen
  \bibfield  {author} {\bibinfo {author} {\bibfnamefont {Nicolas}\ \bibnamefont
  {Boumal}}, \bibinfo {author} {\bibfnamefont {Bamdev}\ \bibnamefont {Mishra}},
  \bibinfo {author} {\bibfnamefont {P.-A.}\ \bibnamefont {Absil}},\ and\
  \bibinfo {author} {\bibfnamefont {Rodolphe}\ \bibnamefont {Sepulchre}},\
  }\bibfield  {title} {\bibinfo {title} {{Manopt, a Matlab Toolbox for
  Optimization on Manifolds}},\ }\href
  {http://jmlr.org/papers/v15/boumal14a.html} {\bibfield  {journal} {\bibinfo
  {journal} {J. Mach. Learn. Res.}\ }\textbf {\bibinfo {volume} {15}},\
  \bibinfo {pages} {1455} (\bibinfo {year} {2014})}\BibitemShut {NoStop}%
\bibitem [{\citenamefont {Li}(2026)}]{our_codes}%
  \BibitemOpen
  \bibfield  {author} {\bibinfo {author} {\bibfnamefont {J.}~\bibnamefont
  {Li}},\ }\href@noop {} {\emph {\bibinfo {title} {{Matlab codes for this
  work}}}} (\bibinfo {year} {2026}),\ \bibinfo {note}
  {https://zenodo.org/records/18183080}\BibitemShut {NoStop}%
\bibitem [{\citenamefont {Chau}\ and\ \citenamefont {Zeng}(2024)}]{CZbound}%
  \BibitemOpen
  \bibfield  {author} {\bibinfo {author} {\bibfnamefont {H.~F.}\ \bibnamefont
  {Chau}}\ and\ \bibinfo {author} {\bibfnamefont {W.}~\bibnamefont {Zeng}},\
  }\bibfield  {title} {\bibinfo {title} {{A unifying quantum speed limit for
  time-independent Hamiltonian evolution}},\ }\href@noop {} {\bibfield
  {journal} {\bibinfo  {journal} {J. Phys. A}\ }\textbf {\bibinfo {volume}
  {87}},\ \bibinfo {pages} {235304} (\bibinfo {year} {2024})}\BibitemShut
  {NoStop}%
\bibitem [{\citenamefont {Chau}(2024)}]{CZbound_prog}%
  \BibitemOpen
  \bibfield  {author} {\bibinfo {author} {\bibfnamefont {H.~F.}\ \bibnamefont
  {Chau}},\ }\href@noop {} {\emph {\bibinfo {title} {Wolfram Community}}}
  (\bibinfo {year} {2024}),\ \bibinfo {note}
  {https://community.wolfram.org/groups/-/m/t/3059079}\BibitemShut {NoStop}%
\bibitem [{\citenamefont {Breuer}\ and\ \citenamefont
  {Petruccione}(2007)}]{BP07}%
  \BibitemOpen
  \bibfield  {author} {\bibinfo {author} {\bibfnamefont {Heinz-Peter}\
  \bibnamefont {Breuer}}\ and\ \bibinfo {author} {\bibfnamefont {Francesco}\
  \bibnamefont {Petruccione}},\ }\href@noop {} {\emph {\bibinfo {title} {The
  Theory Of Open Quantum Systems}}}\ (\bibinfo  {publisher} {OUP},\ \bibinfo
  {address} {Oxford},\ \bibinfo {year} {2007})\ \bibinfo {note}
  {\S4.4.1}\BibitemShut {NoStop}%
\bibitem [{\citenamefont {Peres}(1995)}]{Peres}%
  \BibitemOpen
  \bibfield  {author} {\bibinfo {author} {\bibfnamefont {A.}~\bibnamefont
  {Peres}},\ }\href@noop {} {\emph {\bibinfo {title} {Quantum Theory: Concepts
  And Methods}}}\ (\bibinfo  {publisher} {Kluwer},\ \bibinfo {address}
  {Dordrecht},\ \bibinfo {year} {1995})\ \bibinfo {note}
  {{S}ec.~4.4}\BibitemShut {NoStop}%
\bibitem [{\citenamefont {Ray}\ \emph {et~al.}(2022)\citenamefont {Ray},
  \citenamefont {Ghoshal}, \citenamefont {Pati},\ and\ \citenamefont
  {Sen}}]{Ray22}%
  \BibitemOpen
  \bibfield  {author} {\bibinfo {author} {\bibfnamefont {Tanaya}\ \bibnamefont
  {Ray}}, \bibinfo {author} {\bibfnamefont {Ahana}\ \bibnamefont {Ghoshal}},
  \bibinfo {author} {\bibfnamefont {Arun~Kumar}\ \bibnamefont {Pati}},\ and\
  \bibinfo {author} {\bibfnamefont {Ujjwal}\ \bibnamefont {Sen}},\ }\bibfield
  {title} {\bibinfo {title} {{Estimating quantum coherence by noncommutativity
  of any observable and its incoherent part}},\ }\href
  {https://doi.org/10.1103/PhysRevA.105.062423} {\bibfield  {journal} {\bibinfo
   {journal} {Phys. Rev. A}\ }\textbf {\bibinfo {volume} {105}},\ \bibinfo
  {pages} {062423} (\bibinfo {year} {2022})}\BibitemShut {NoStop}%
\bibitem [{\citenamefont {Apostol}(1974)}]{Apostol}%
  \BibitemOpen
  \bibfield  {author} {\bibinfo {author} {\bibfnamefont {Tom~M.}\ \bibnamefont
  {Apostol}},\ }\href@noop {} {\emph {\bibinfo {title} {{Mathematical
  Analysis}}}},\ \bibinfo {edition} {2nd}\ ed.\ (\bibinfo  {publisher}
  {Addison-Wesley},\ \bibinfo {address} {Reading, MA},\ \bibinfo {year}
  {1974})\BibitemShut {NoStop}%
\end{thebibliography}%
	
\end{document}